\documentclass[usenatbib]{mn2e}
\usepackage{epsfig}
\usepackage{natbib}
\usepackage{pdfsync}

\newcommand{\etal}{et~al.} 
\newcommand{\ionhy}{H{\sc ii} }

\newcommand{\kms}{$\mbox{km~s}^{\scriptsize{-1}}$}

\begin{document}

\title[19.9-GHz methanol masers]{The first high-resolution observations of 19.9-GHz methanol masers}
\author[Krishnan \etal]{V. Krishnan,$^{1,2}$\thanks{Email: Vasaant.Krishnan@utas.edu.au}, S.\ P. Ellingsen,$^{1}$ M.\ A. Voronkov,$^{2,1}$ S.\ L. Breen$^{2}$\\
  \\
  $^1$ School of Mathematics and Physics, University of Tasmania, Private Bag 37, Hobart, Tasmania 7001, Australia\\
  $^2$ CSIRO Astronomy and Space Science, Australia Telescope National Facility, CSIRO, PO Box 76, Epping, NSW 1710, Australia}

 \maketitle

\begin{abstract}
We have used the Australia Telescope Compact Array to undertake the first high-resolution observations of the 19.9-GHz (2$_1-3_0$E) methanol maser transition.  The emission is coincident with the location of the targeted 6.7-GHz methanol masers to within 0.2 arcseconds in absolute position.  We find that the relative distribution of the 19.9-GHz maser emission in our sample differs from that observed in the 6.7-GHz transition, in contrast to the similar distribution often seen in the 6.7- and 12.2-GHz methanol masers in many sources.  We also find that the peak velocity for the 19.9-GHz methanol masers is frequently different from that observed in other class~II transitions in the same regions.  These two results suggest that while 19.9-GHz methanol masers arise from the same general location as other class~II transitions, they are likely not coincident on milliarcsecond scales; which has implications for multi-transitional modelling of class~II methanol maser sources.  We have investigated the properties of the OH and other class~II methanol transitions for those 107-GHz methanol masers with and without an associated 19.9-GHz methanol maser.  On the basis of these comparisons we suggest that the 19.9-GHz transition traces an evolutionary phase around the peak of the class~II maser luminosity into its decline.
\end{abstract}


\section{Introduction}
Water, OH and methanol molecules commonly show maser emission towards young, high-mass star formation regions.  Amongst these, methanol has the richest spectrum in the radio and millimetre regime and shows the largest number (in excess of 30) of different maser transitions.  The methanol maser transitions are empirically classified into two classes \citep{Menten91a,Batrla+87}.  The class~I maser transitions are collisionally pumped and are associated either with molecular outflows \citep{Voronkov+06,Cyganowski+09}, or other shocks \citep{Voronkov+10a} within $\sim$ 1~pc of a young high-mass star \citep{Kurtz+04}.  The best studied class~I methanol masers are the 44.1-GHz \citep[e.g.][]{Cyganowski+09} and the 95.1-GHz \citep[e.g.][]{Chen+11} transitions.  The class~II methanol masers are found close to very young high-mass stars \citep[e.g.][]{Ellingsen06} and are often associated with ground-state OH masers \citep{Caswell+95a} and water masers \citep{Szymczak+05}.  The best studied class~II methanol maser transitions are the 6.7- and 12.2-GHz.  To date more than 900 class~II methanol maser regions have been identified in the Galaxy, with a sensitive, complete survey of the southern sky recently completed in the 6.7-GHz transition \citep{Caswell+10,Caswell+11,Green+10,Green+12a}.  In this paper we focus solely upon class~II methanol masers, of which a total of 18 different transitions, or transition groups have been identified to date \citep[see][and references therein]{Ellingsen+12}.

In total twenty--nine methanol maser sites have been searched for the class~II 19.9-GHz methanol masers \citep{Wilson+85,Menten+89a,Ellingsen+04}, resulting in nine detections. The search undertaken by \citet{Ellingsen+04} using the Tidbinbilla 70-m antenna was significantly more sensitive than most searches for rare, weak methanol maser transitions, and resulted in the detection of a number of sources with peak flux density less than 0.5~Jy.  \citet{Ellingsen+04} also found that the 19.9-GHz methanol masers appeared to be preferentially associated with class~II maser sites with an ultra-compact \ionhy region and 6035-MHz OH masers. Seven of the nine known 19.9-GHz methanol maser sources are visible from the southern hemisphere, the two exceptions are W3(OH) and NGC7538.

The presence of an interstellar maser in a region signposts the presence of physical conditions capable of producing population inversion for that transition \citep[e.g.][]{Cragg+05}. However, the intensity of the observed maser emission depends (non-linearly) on a number of factors, many of which are poorly constrained observationally, and so inferring the specific physical conditions in the maser region from observations of its intensity is in general not possible. Where comparative observations have been made of the 6.7- and 12.2-GHz methanol masers they show that emission at the same velocity is spatially coincident on scales of tens of milliarcseconds, or less \citep{Menten+92,Norris+93,Minier+00}. If there are multiple maser transitions observed from the same molecule and in the same general vicinity (sub-arcsecond scales), theoretical models are much better constrained as they are then required to reproduce both the presence of the different maser transitions, as well as the observed intensity ratio. 

Multi-transition modelling using methanol masers has been undertaken in just a few sources which show emission in a larger number of transitions. Examples of such sources include W3(OH) \citep{Sutton+01}, NGC6334F and G\,$345.01+1.79$ \citep{Cragg+01}.  These studies used some low angular resolution (single-dish) data, and assume that where emission is observed at the same velocity it originates from the same region, in terms of spatial location and dimensions. The first of these assumptions is likely to be correct, as it has been explicitly demonstrated for the two strongest and most common class~II methanol maser transitions.  The second assumption: that the maser ``spot'' size is the same at all frequencies is not likely to be correct, as differences have been shown in studies of spatially coincident 6.7- and 12.2-GHz emission \citep{Minier+02a}, however, in the absence of milliarcsecond-scale observations of all the transitions, it is the simplest one to make. The multi-transitional modelling of \citet{Sutton+01} and \citet{Cragg+01} find conditions in the masing region consistent with those expected in young, high-mass star formation regions, with gas temperatures of around 30K, dust temperatures of around 175K and densities of 10$^6$ cm$^{-3}$.

While the assumption that emission from the different class II methanol maser transitions at the same velocity are spatially coincident seems to be generally true for 6.7- and 12.2-GHz methanol masers, high resolution studies of the less common class II masers are needed to determine if this property extends to other class II transitions. Here we present the first high-resolution observations of the 19.9-GHz methanol maser transition, which will allow us to determine the degree to which its emission is coincident with that observed in the 6.7-GHz (and 12.2-GHz) methanol maser transitions in the same regions.
 
\section{Observations and data reduction}
\label{sec:observe}
\label{sec:observations}
We have used the ATCA in the 6A configuration (baselines ranging from 337 to 5939m) to conduct high resolution 19.9-GHz observations towards nine southern class II methanol maser sites. The sample includes the seven sources detected in the 19.9-GHz transition by \citet{Ellingsen+04} and two additional sources (G\,309.921+0.479 \& G\,$330.953-0.182$) which were detected in later Tidbinbilla observations (unpublished).  For each source we also observed the 6.7-GHz transition, and all measurements were made during a single session in 2005 March 27 (see Table~\ref{tab:coords} for details). The 19.9-GHz (2$_1-3_0$E) transition observations typically consisted of either six or seven 5-minute scans on each target source over a range of hour angles.  In some cases up to nine 5-minute observations were undertaken and the total time onsource for each target is given in Table~\ref{tab:coords}. Each target-source observation was both preceded and followed by a 90-second scan on a phase calibrator, offset on the sky by between 2 -- 12 degrees. For NGC6334F, observations were also made of the 23.1-GHz ($9_2-10_1$A$^{+}$) transition of methanol. A total of six 5-minute scans of this source were made at 23.1~GHz, again over a range of hour angles and in between phase calibrator scans. These 23.1-GHz observations were undertaken after each 19.9-GHz scan of NGC6334F. The 6.7-GHz ($5_1-6_0A^+$) transition observations consisted of a total of three (or more) 90-second scans over a range of hour angles, preceded and followed by 90-second observations of the same phase calibrators as used in the 19.9-GHz observations. Table~\ref{tab:coords} gives the pointing centre for the observations, the total time on source for each transition and the name of the phase calibrator. Observations of the different maser transitions were interleaved to minimise the potential sources of error, in the absolute position of measurements at two different frequencies. A discussion of the observed difference in the absolute position of the different maser transitions is presented in Section \ref{sec:results}.

PKS\,B1253$-$055 was observed as the bandpass calibrator for the 19.9- and 23.1-GHz observations. Primary flux density calibration is relative to PKS\,B1934$-$638 which was also used as the bandpass calibrator for the 6.7-GHz observations. The assumed flux densities for PKS\,B1934-638 were 3.92~Jy, 0.93~Jy and 0.80~Jy for the 6.7-, 19.9- and 23.1-GHz transitions respectively and based on ATNF MIRIAD's built--in flux scale \citep{Sault+03, Reynolds+94}. The absolute flux density calibration is expected to be accurate to better than 10 percent.

The correlator was set to record 1024 spectral channels for both parallel- and cross-polarization products across a 4-MHz bandwidth for all frequencies. This corresponds to a velocity coverage of 179.8, 60.1 and 51.9~\kms , and channel separations of 0.176, 0.058 and 0.051~\kms ~at 6.7, 19.9~ and 23.1~GHz respectively. With uniform weighting of the autocorrelation function, this corresponds to velocity resolutions of 0.211, 0.070 and 0.061~\kms\/.

Data reduction was performed using the MIRIAD package using the built-in model to correct for the differential opacity between the source and calibrators due to their slightly different elevations. Channels associated with line emission were deconvolved, and the spectra of the masers extracted (shown in Figure \ref{fig:spectrum} and with details in Table \ref{tab:maserList}). These were then imaged to determine the spatial distribution of the emission (Figures \ref{fig:w309} to \ref{fig:w353}). The line-free spectral channels were averaged together before imaging to investigate the continuum emission associated with each of the maser targets (Figures \ref{fig:cont309} to \ref{fig:cont353} and Table \ref{tab:contList}). We adopted rest frequencies of 6.6685192(8), 19.9673961(2) and 23.1210242(5)~GHz (numbers in brackets indicate the magnitude of uncertainty in the last digit) for the three observed methanol maser transitions \citep{Muller+04}. The uncertainty in the velocity scale due to the accuracy to which the rest frequencies have been determined are 0.036, 0.003 and 0.007~\kms\/, at 6.7, 19.9 and 23.1~GHz respectively (i.e. much less than the spectral resolution in each case). The synthesised beam size for the observations were approximately 0.6 $ \times $ 0.4 arcsec$ ^2 $ for sources at 19.9 and 23.1~GHz and 2.2 $ \times $ 1.6 arcsec$ ^2 $ for those at 6.7~GHz.

The absolute position of each maser transition was determined by imaging the velocity channel corresponding to the strongest emission at 19.9-GHz. We then fitted a 2D Gaussian, with the same dimensions as the synthesised beam, to the emission in the deconvolved image to determine the absolute position of the peak maser emission associated with the source. Once the absolute position of each maser source had been determined, we self-calibrated the strongest emission channel for the brighter sources and then applied these solutions to all other spectral channels for that source. The resulting improvement can be seen from comparison of the RMS noise of the image frames between those sources which have and have not been self-calibrated (Table \ref{tab:maserList}). The RMS noise in the image of an individual spectral channel at 19.9-GHz if self-calibration was performed is typically $<$10 mJy. The 19.9-GHz maser emission in G\,$328.808+0.633$ and G\,$345.003-0.224$ was too weak for reliable self-calibration and the associated image RMS for these sources is approximately a factor of 5 times greater. The RMS itself was measured from a square box of $\sim 10$ arcsec$^2$ in a deconvolved image frame which was free from any maser emission and away from the pointing centre. Either one or two iterations of self-calibration (phase then amplitude), was undertaken using a one minute solution interval, using the clean components as the initial model. The use of self-calibration enabled us to more accurately investigate the relative distribution of the maser features within an individual source.

\begin{table*}
\centering
\caption{Observation details. Target positions are based on the 6.7-GHz positions from \citet{Caswell+95a, Caswell+95b}. The phase calibrator for NGC6334F at the 23.1-GHz observation was 1759$-$39.}
\begin{tabular}{lllcccc}\hline
 \multicolumn{1}{c}{\bf Source} & \multicolumn{1}{c}{\bf RA} & \multicolumn{1}{c}{\bf Dec} &\multicolumn{3}{c}{\bf Observation} \\
 \multicolumn{1}{c}{\bf Name} & \multicolumn{1}{c}{\bf (J2000)} & \multicolumn{1}{c}{\bf (J2000)}& \multicolumn{3}{c}{\bf Duration (mins)} & \multicolumn{1}{c}{\bf Phase} \\ 
            &              &            & & && \multicolumn{1}{c}{\bf Calibrator} \\
  & \multicolumn{1}{c}{\bf $h$~~~$m$~~~$s$} & \multicolumn{1}{c}{\bf $^\circ$~~~$\prime $~~~$\prime \prime $} & \multicolumn{1}{c}{\bf 19.9~GHz} & \multicolumn{1}{c}{\bf 6.7~GHz} & \multicolumn{1}{c}{\bf 23.1~GHz}\\
\hline
\hline
 G\,$309.921+0.479$ & 13 50 41.77  & $-$61 35 10.1   & 45  & 7 &       &1352$-$63 \\
 G\,$323.740-0.263$  & 15 31 45.45  & $-$56 30 50.1   & 40  & 7 &       &1511$-$55 \\
 G\,$328.808+0.633$ & 15 55 48.45  & $-$52 43 06.6   & 35  & 7 &       &1511$-$55 \\
 G\,$330.953-0.182$  & 16 09 52.83  & $-$51 54 57.6   & 35  & 7 &       &1613$-$586 \\
 G\,$339.884-1.259$  & 16 52 04.66  & $-$46 08 34.2   & 30  & 6 &       &1646$-$50 \\
 G\,$345.003-0.223$  & 17 05 10.89  & $-$41 29 06.2   & 35  & 5 &       &1729$-$37 \\
 G\,$345.010+1.792$ & 16 56 47.58  & $-$40 14 25.8   & 35  & 5  &      &1729$-$37 \\
 NGC6334F                  & 17 20 53.37  & $-$35 47 01.2   & 30  & 3  & 30 &1729$-$37\\
 G\,$353.410-0.360$  & 17 30 26.18  & $-$34 41 45.6   & 34  & 4  &      &1759$-$39 \\
\hline
\end{tabular}
  \label{tab:coords}
\end{table*}

\section{Results}
\label{sec:results}

\begin{figure*}
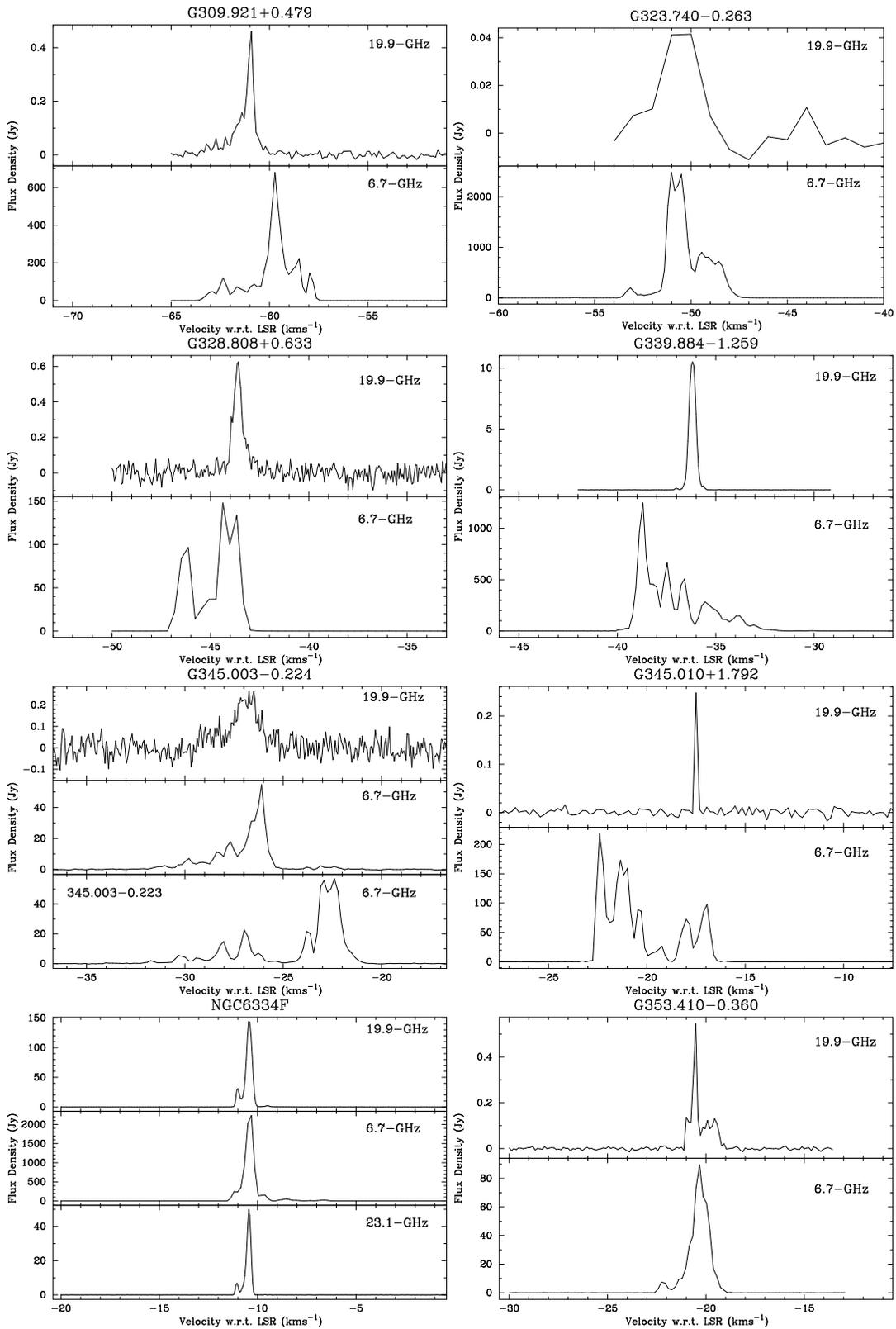

   \begin{center}
   \begin{minipage}[c]{0.4\textwidth}
       \psfig{file=309_spectra.eps,angle=-90,width=\textwidth}
   \end{minipage}
   \begin{minipage}[c]{0.4\textwidth}
       \psfig{file=323_spectra.eps,angle=-90,width=\textwidth}
   \end{minipage}
   \begin{minipage}[c]{0.4\textwidth}
       \psfig{file=328_spectra.eps,angle=-90,width=\textwidth}
   \end{minipage}
   \begin{minipage}[c]{0.4\textwidth}
       \psfig{file=339_spectra.eps,angle=-90,width=\textwidth}
   \end{minipage}
   \begin{minipage}[c]{0.4\textwidth}
       \psfig{file=345_spectra.eps,angle=-90,width=\textwidth}
   \end{minipage}
   \begin{minipage}[c]{0.4\textwidth}
       \psfig{file=345010_spectra.eps,angle=-90,width=\textwidth}
   \end{minipage}
   \begin{minipage}[c]{0.4\textwidth}
       \psfig{file=ngc_spectra.eps,angle=-90,width=\textwidth}
   \end{minipage}
   \begin{minipage}[c]{0.4\textwidth}
       \psfig{file=353_spectra.eps,angle=-90,width=\textwidth}
   \end{minipage}
 \end{center}
\caption{Spectra of the 19.9-, 6.7- and 23.1-GHz methanol masers extracted from ATCA image cubes. G\,$345.003-0.224$ is presented in conjunction with G\,$345.003-0.223$ to show the extent of the maser emission overlap. See Section \ref{sec:subsec345} for detailed comments on each source.}
  \label{fig:spectrum}
\end{figure*}

The spectra of all observed 6.7-, 19.9- and 23.1-GHz methanol masers (presented in Figure \ref{fig:spectrum}) have been extracted from regions of the image cube approximately the size of the synthesised beam (0.6$^{\prime\prime} $ at 19.9- and 23.1-GHz, and  2.0$^{\prime \prime} $ at 6.7-GHz), centred at the location of the strongest emission in the cube. In general, the 19.9-GHz methanol masers contrast with the 6.7-GHz masers which are much stronger (peak flux density $>100$~Jy) and cover a much larger velocity range. These 19.9-GHz methanol masers are in many cases weak ($<1$~Jy) and often only have a single spectral feature. A detailed comparison between the spectra at both frequencies is made in Section \ref{sec:detailedComments} on a source-by-source basis.

Table \ref{tab:nh3Compare} lists the thermal velocities and angular separation of the centroid of NH$_3$ \citep{Purcell+12} and CS(2--1) \citep{Bronfman+96} thermal emission associated with the observed methanol masers. The only exception is G\,$345.010+1.792$, which was not observed in either of those studies. However, \citet{Ellingsen+12} measured a thermal velocity of $-$12.8~\kms\/ at 216.9 GHz for this source, compared to peak maser LSR velocities at $-17.5$ \kms\/ at 19.9 GHz and $-22.4$ \kms\/ at 6.7 GHz (Table \ref{tab:maserList}). The peak velocity of the methanol maser emission at 19.9 and 6.7 GHz in this study are within 5~\kms\/ of the thermal velocities of the cloud in all cases but for G\,$339.884-1.259$, where the difference is of magnitude 7.2 \kms. This is consistent with studies undertaken by \citet{Xu+09,Pandian+09}, and shows that the maser peak velocity is a good indicator of the systemic velocities of the regions. \citet{Pandian+09} showed that the median LSR velocity of the maser is usually close to the systemic velocities of the regions and so we present these in Table \ref{tab:nh3Compare} for the 6.7-GHz emission.

\begin{table*}
\centering
\caption{A comparison between the 6.7- and 19.9-GHz methanol maser peak velocity and the velocity centroids of the NH$_3$ \citep{Purcell+12} and CS(2--1) \citep{Bronfman+96} emission associated with the same star formation region. The separation columns show the proximity of the centroid of the NH$_3$ or CS cloud to the absolute position of the peak of the 19.9-GHz maser as presented in Table \ref{tab:maserList}. \citet{Purcell+12} report emission from the ammonia NH$_3$ (1,1) transition for G\,$309.921+0.479$ and G\,$323.740-0.263$, and the NH$_3$ (2,2) transition for G\,$345.003-0.224$ and G\,$353.410-0.360$.}
\begin{tabular}{lccccccc}\hline
 \multicolumn{1}{c}{\bf Source} & \multicolumn{2}{c}{\bf NH$_3$}  & \multicolumn{2}{c}{\bf CS(2--1)} &\multicolumn{1}{c}{\bf 19.9-GHz} &\multicolumn{1}{c}{\bf 6.7-GHz}&\multicolumn{1}{c}{\bf 6.7-GHz}\\
 \multicolumn{1}{c}{\bf Name}  & \multicolumn{1}{c}{\bf Centroid Velocity}& \multicolumn{1}{c}{\bf Separation } &\multicolumn{1}{c}{\bf Centroid Velocity} & \multicolumn{1}{c}{\bf Separation} & \multicolumn{2}{c}{\bf Peak velocity}& \multicolumn{1}{c}{\bf Median velocity}\\
& \multicolumn{1}{c}{\bf (\kms)} & \multicolumn{1}{c}{\bf($^{\prime\prime}$)} &\multicolumn{1}{c}{\bf (\kms)} & \multicolumn{1}{c}{\bf($^{\prime\prime}$)} & \multicolumn{2}{c}{\bf (\kms)}& \multicolumn{1}{c}{\bf (\kms)}\\
\hline
\hline
 G\,$309.921+0.479$ & $-$57.7  & 26.2 & $-$58.4 & 2.0   &$-$60.9 & $-$59.7 & $-$60.8\\
 G\,$323.740-0.263$  & $-$50.5  & 50.4 & $-$49.5 & 31.1 &$-$50.0 & $-$51.0 & $-$60.0\\
 G\,$328.808+0.633$ & $-$         & $-$  & $-$42.3 & 3.7   &$-$43.6 & $-$44.4  & $-$45.1\\
 G\,$339.884-1.259$  & $-$         & $-$  & $-$31.6 & 14.1 &$-$36.2 & $-$38.7  & $-$35.0\\
 G\,$345.003-0.224$  & $-$28.2  & 9.9   & $-$22.2 & 18.1 &$-$26.5 & $-$26.1  & $-$26.9\\
 NGC6334F                    &       $-$   & $-$  & $-$6.9   & 17.2 &$-$10.5 & $-$10.3  & $-$9.0\\
 G\,$353.410-0.360$  & $-$17.8  & 45.6 & $-$16.7 & 33.4 &$-$20.5 & $-$20.3  & $-$20.8\\
\hline
\end{tabular}
  \label{tab:nh3Compare}
\end{table*}

The (unpublished) Tidbinbilla detection of G\,$330.953-0.182$ was marginal and the current ATCA observations detected no emission at 19.9-GHz towards G\,$330.953-0.182$ above a limit of 90 mJy which corresponds to 3 times the RMS noise. For the remaining sources we were able to detect maser emission in one or more individual spectral channels (velocity width 0.070~\kms), the only exception was G\,$323.740-0.263$, for which it was necessary to average four spectral channels to obtain sufficient signal to noise to detect 19.9-GHz emission. Spectral averaging was also employed for G\,$345.010+1.792$, and even though self-calibration was not employed in imaging either of these sources, the RMS in the image frames was measured to be $<$10 mJy as a result of averaging.

We have compared our ATCA spectra to those of \citet{Ellingsen+04} (observations made October-November 2003 with the Tidbinbilla 70-m antenna), and find that the 19.9-GHz spectra have in general changed very little in the 16 month time interval between the two sets of observations.  The only exception is again G\,$323.740-0.263$, which is weaker by a factor of 4 in the current observations. Where changes are observed, they are discussed in more detail, on a source-by-source basis in Section \ref{sec:detailedComments}.

We examined the intensity of the strongest maser emission for each source as a function of baseline length and found it to be consistent with the masers being unresolved for a 6-km baseline, implying that the actual angular size of the 19.9-GHz emission is less than 0.6$^{\prime\prime}$ (the approximate dimensions of the synthesised beam). A small decrease in flux density was observed, but this appears to be due to slight decorrelation on the longest baselines, rather than source structure. If we assume that the emission for all sources and all transitions was unresolved we are able to use the peak flux density and the synthesised beam size to calculate a lower limit on the brightness temperature of the emission in each case (see last column in Table~\ref{tab:maserList}). With the exception of G\,$323.740-0.263$, the brightness temperature exceeds $\sim$3000~K for each of the sources.

One of the primary aims of these observations was to determine if the methanol maser emission at 6.7 and 19.9~GHz is spatially coincident.  This was achieved by first identifying the single 6.7-GHz spectral channel with velocity closest to that of the 19.9-GHz methanol maser peak velocity, and measuring the difference in the absolute position between the two frequencies at those identified velocities. We then undertook the same measurement between the position of the emission in all other spectral channels at 6.7~GHz, with the same position of the peak maser emission at 19.9~GHz. Our findings showed that for all sources the minimum difference in position occurs for the 6.7~GHz velocity channel most closely corresponding to the velocity at the 19.9-GHz maser peak. This difference was measured to be between 0.1 -- 0.35$^{\prime \prime }$, with the angular and corresponding linear offsets summarised in Table~\ref{tab:sizeDist} and the third last column of the 19.9-GHz masers' rows of Table \ref{tab:maserList}.  A detailed analysis of positional coincidence between the primary and any secondary maser components between the two transitions is presented in Section \ref{sec:relativePos}.

The mean offset in absolute position between the 19.9- and 6.7-GHz methanol maser emission at the same velocity was calculated to be 0.2$^{\prime \prime }$. We compared this to the absolute positions that were determined for the spectral channel containing the 6.7-GHz methanol maser peak emission with those reported in the Methanol Multi-Beam survey \citep{Green+10,Green+12a,Caswell+10}. We find the RMS difference between the positions of the same 6.7-GHz methanol masers observed at different epochs (with the same instrument) to be 0.4$^{\prime \prime }$. The absolute positions of the 6.7-GHz masers in \citet{Green+10,Green+12a,Caswell+10} were determined from frames with the peak flux emission for each source. The formal errors reported in the fitting process for the positions of the masers are $\sim$ 0.03$^{\prime \prime }$, however, while this is likely to be an underestimate the uncertainty in absolute positions due to the accuracy of the calibration will dominate any contribution due to fitting Gaussian profiles to the maser images.  A comparison of the LSR velocities of the peak of the spectra of the 6.7-GHz masers, between the Methanol Multi-Beam survey and this study, shows that identified peaks are always similar to $\leq$0.5 \kms. This means that positional measurements between the different sets of observations were with respect to the same spectral feature. The 0.4$^{\prime \prime }$ RMS result between the positional differences between the different sets of 6.7-GHz observations corresponds to the expected positional accuracy for ATCA observations made in good weather, with unresolved phase calibrators \citep{Caswell97}. We attribute the closer correspondence between our 19.9- and 6.7-GHz measurements made on the same day, and that between our 6.7-GHz positions compared to \citet{Green+10,Green+12a,Caswell+10} to the observation strategy (interleaved observations of the two transitions using the same phase calibrators). We were able to test this assertion through the use of additional phase calibrators which were observed for G\,$328.808+0.633$,  G\,$339.884-1.259$ and NGC6334F at 6.7 and 19.9 GHz. For each of these additional calibrators, the channels containing continuum emission were averaged, CLEANed and imaged, using the primary phase calibrators (Table \ref{tab:coords}), while following standard reduction procedures. The absolute positions of the additional calibrators were then determined using a 2D Gaussian fit, and the RMS between these at 6.7 and 19.9 GHz was better than 0.2$^{\prime \prime }$.

Our observations also show that six of the eight 19.9-GHz methanol masers are projected against radio continuum with a peak flux density in excess of 100~mJy at 20~GHz, and we present the details in Section \ref{sec:maserContinuum}.

\begin{table*}
\centering
\caption{A summary of the 19.9-, 6.7- and 23.1-GHz methanol maser characteristics. All velocities are given with respect to the local standard of rest (LSR) frame.  V$_{\mbox{\scriptsize{Low}}}$ and  V$_{\mbox{\scriptsize{High}}}$ give the velocity range for which maser emission is observed for each source above 5 times the RMS. The offset for the 6.7-GHz observations is a comparison of the coordinates of the peak of the emission determined from the current observations, and those reported by \citet{Green+10,Green+12a,Caswell+10}. The offset listed for the 19.9- and 23.1-GHz masers is the difference with respect to the 6.7-GHz emission from the current observations at the velocity of 19.9-GHz maser peak. The final column gives a lower limit to the brightness temperature of the sources observed at 19.9~GHz, calculated from the peak flux density assuming an upper limit to the angular size of the source of 0.62 arcsec (typical synthesised beam dimension).}
\begin{tabular}{llllllllccccc}\hline
& \multicolumn{1}{c}{\bf Source}& \multicolumn{1}{c}{\bf RA} & \multicolumn{1}{c}{\bf Dec} & \\
& \multicolumn{1}{c}{\bf Name} & \multicolumn{1}{c}{\bf (J2000)} & \multicolumn{1}{c}{\bf (J2000)} & & \\ 
&            &              &            & \bf Peak & \bf Velocity & \bf V$_{\mbox{\scriptsize{Low}}}$ & \bf V$_{\mbox{\scriptsize{High}}}$ & \multicolumn{1}{c}{\bf Offset} &  \multicolumn{1}{c}{\bf RMS}&  $T_B \times 10^3 $  &  \\
&  & \multicolumn{1}{c}{\bf $h$~~~$m$~~~$s$} & \multicolumn{1}{c}{\bf $^\circ$~~~$\prime $~~~$\prime \prime $} & \bf (Jy) & \multicolumn{3}{c}{\bf (\kms)}& \multicolumn{1}{c}{\bf ($\prime \prime $)}& \multicolumn{1}{c}{\bf (mJy)}& \multicolumn{1}{c}{\bf (K)}  & & \\
\hline
\hline
\bf 19.9~GHz & \\
& G\,$309.921+0.479$     & 13 50 41.78 & $-$61 35 10.8  & 0.46   & $-$60.9  & $-$61.9 & $-$60.4 & 0.17 & 8   & 5.4\\
& G\,$323.740-0.263$     & 15 31 45.39 & $-$56 30 50.5  & 0.04   & $-$50.0  & $-$52.0 & $-$50.0 & 0.25  & 7   & 0.5\\
& G\,$328.808+0.633$     & 15 55 48.45 & $-$52 43 06.8  & 0.63   & $-$43.6  & $-$44.0 & $-$43.1 &  0.29 &40 & 7.3\\
& G\,$339.884-1.259$     & 16 52 04.60 & $-$46 08 34.1  & 10.5   & $-$36.2  & $-$37.1 & $-$35.5 & 0.23  & 9   & 122.\\
& G\,$345.003-0.224$     & 17 05 11.21 & $-$41 29 07.0  & 0.26   & $-$26.5  & $-$27.3 & $-$26.4 & 0.10  & 40 & 3.0\\
& G\,$345.010+1.792$    & 16 56 47.53 & $-$40 14 25.8  & 0.25   & $-$17.5  & $-$17.7 & $-$17.3 & 0.34  & 7    & 2.9\\
& NGC6334F                     & 17 20 53.36 & $-$35 47 01.5  & 144.   & $-$10.5  & $-$11.4 & $-$8.7   & 0.10  & 20 & 1870.\\
& G\,$353.410-0.360$     & 17 30 26.17 & $-$34 41 45.6  & 0.55   & $-$20.5  & $-$21.1 & $-$19.3 & 0.10  & 6   & 6.3\\
\bf 23.1~GHz                    & \\
& NGC6334F                     & 17 20 53.35 & $-$35 47 01.8  & 50.0    & $-$10.5  & $-$11.2 & $-$10.2  & 0.23 & 30\\
\bf 6.7~GHz                     & \\
& G\,$309.921+0.479$    & 13 50 41.76 & $-$61 35 10.7  & 681.    & $-$59.7  & $-$64.1 & $-$57.6 & 0.52 & 30 \\
& G\,$323.740-0.263$     & 15 31 45.42 & $-$56 30 50.5  & 2490.  & $-$51.0  & $-$55.1 & $-$46.8 & 0.47 & 40\\
& G\,$328.808+0.633$    & 15 55 48.42 & $-$52 43 06.7  & 148.    & $-$44.4  & $-$46.8 & $-$43.3 & 0.29 & 30 \\
& G\,$330.953-0.182$     & 16 09 52.35 & $-$51 54 57.2  & 5.7      & $-$87.7  & $-$88.9 & $-$87.3 & 0.44 & 60\\
& G\,$339.884-1.259$     & 16 52 04.69 & $-$46 08 34.5  & 1250.  & $-$38.7  & $-$40.3 & $-$29.6 & 0.36 & 40\\
& G\,$345.003-0.224$     & 17 05 11.21 & $-$41 29 06.9  & 54.7    & $-$26.1  & $-$31.6 & $-$22.2 & 0.22 & 200 \\
& G\,$345.010+1.792$    & 16 56 47.56 & $-$40 14 25.8  & 218.    & $-$22.4  & $-$22.9 & $-$16.6 & 0.23 & 40 \\
& NGC6334F                     & 17 20 53.36 & $-$35 47 01.6  & 2230.  & $-$10.3  & $-$11.4 & $-$6.5   & 0.42 & 1000 \\
& G\,$353.410-0.360$     & 17 30 26.17 & $-$34 41 45.7  & 89.6    & $-$20.3  & $-$22.4 & $-$19.1 & 0.16 & 50 \\
\hline
\end{tabular}
  \label{tab:maserList}
\end{table*}

\subsection{Relative positions}
\label{sec:relativePos}

The high spectral resolution of the 19.9- and 23.1-GHz observations means that around 5 channels cover the width of a single maser spectral feature. We also find that the 19.9- and 23.1-GHz emission mostly consists of 2 or 3 maser components. In contrast, despite the coarser spectral resolution at 6.7 GHz, we typically identified twice as many maser components for this transition.

All of the 6.7-GHz methanol masers observed in this project have previously been imaged with the ATCA either by \citet{Norris+93}, \citet{Ellingsen+96a}, \citet{Phillips+98b}, \citet{Walsh+98} or \citet{Caswell97}.  Their observations were more sensitive than those undertaken here, as we spent the majority of observing time on the weaker 19.9-GHz maser emission.  The correlator configuration was also selected to be optimal for the 19.9-GHz observations, and the main purpose for determining the distribution of the 6.7-GHz emission here is to enable a direct single epoch comparison with the 19.9-GHz masers, without any uncertainty regarding temporal variability.

We infer that the regions responsible for the maser emission from the two transitions are coincident to within the accuracy of these observations (approximately 0.2$^{\prime\prime}$), as described in Section \ref{sec:results}.  This is also comparable to the typical size of class~II methanol maser clusters \citep{Caswell97}. Table \ref{tab:sizeDist} summarises the distances and separation associated with the masers and their constituent components as identified in Figures \ref{fig:w309} to \ref{fig:w353}. The detailed morphology of the maser components for each source is discussed further in Section \ref{sec:detailedComments}.

\begin{table*}
\centering
\caption{Estimates of the separation between individual maser components identified at 6.7 and 19.9 GHz. The values presented in the 19.9-GHz and 6.7-GHz columns represent the maximum separation between the maser components (where there is more than a single component) for that transition in arcseconds and milliparsecs. The offsets have been taken from Table \ref{tab:maserList} and describe the difference in absolute position between the peak emission at 19.9~GHz and the emission at the corresponding velocity at 6.7-GHz. Distances have been adopted from \citet{Green+11}$^1$; \citet{Caswell+11a}$^2$; \citet{Ellingsen+96a}$^3$. The two maser components at 23.1 GHz for NGC6334F are separated by 0.1 arcsec with an offset of 0.23 arcsec from the position of the 19.9-GHz peak.}
\begin{tabular}{lccccccc}\hline
\multicolumn{1}{c}{\bf Source} & \\
\multicolumn{1}{c}{\bf Name} & \multicolumn{2}{c}{\bf 19.9~GHz} & \multicolumn{2}{c}{\bf 6.7~GHz} & \multicolumn{2}{c}{\bf Offset}  & \multicolumn{1}{c}{\bf Distance}\\
 & \multicolumn{1}{c}{\bf ($^{\prime\prime}$)}  &\multicolumn{1}{c}{\bf (mpc)} &\multicolumn{1}{c}{\bf ($^{\prime\prime}$)}  & \multicolumn{1}{c}{\bf (mpc)} & \multicolumn{1}{c}{\bf ($^{\prime\prime}$)} & \multicolumn{1}{c}{\bf (mpc)} &\multicolumn{1}{c}{\bf (kpc)}\\
\hline
\hline
 G\,$309.921+0.479$     & 0.2 & 2.3  & 0.8 & 9.3 & 0.17 & 2.0 & 4.8$^1$ \\
 G\,$323.740-0.263$      & $-$& $-$ & 0.2 & 1.4 & 0.25 & 1.7 & 2.8$^1$ \\
 G\,$328.808+0.633$     & 0.2 & 1.3  & 0.5 & 3.2 & 0.29 & 1.8 & 2.6$^1$ \\
 G\,$339.884-1.259$      & 0.1 & 0.6  & 0.8 & 5.0 & 0.23 & 1.4 & 2.6$^1$ \\
 G\,$345.003-0.224$      & $-$& $-$ & 1.0 & 6.5 & 0.10 & 0.7 & 2.7$^1$ \\
 G\,$345.010+1.792$     & $-$& $-$ & 0.2 & 1.0 & 0.34 & 0.6 & 2.0$^1$ \\
 NGC6334F                     & 0.2 & 0.8  & 4.0 & 16. & 0.10 & 0.4 & 1.7$^3$ \\
 G\,$353.410-0.360$      & 0.1 & 1.1  & 0.1 & 1.1 & 0.10 & 1.1 & 4.5$^2$ \\
\hline
\end{tabular}
  \label{tab:sizeDist}
\end{table*}

\subsubsection{Comments on individual sources}
\label{sec:detailedComments}
Figures \ref{fig:w309} to \ref{fig:w353} show the individual masers components which were identified from the spectral image cubes of the 6.7-, 19.9- and 23.1-GHz observations. The spectral channels with weaker maser emission in the image cubes had lower signal-to-noise and correspondingly higher uncertainty in position. In these cases, it was important to distinguish between real maser emission and residual noise in the image. We therefore used a flux density weighted algorithm to group the relevant frames with associated emission into individual maser components, thus ensuring that positional and velocity information from the strongest masers are accurately represented. Our 6.7-GHz results indicate the efficacy of this technique, as they can be directly compared with previous observations of the same sources which were made with higher spectral resolution and sensitivity (thus representing superior data on the number and relative distribution of these masers). In cases where there was only a single 19.9-GHz maser component which was identifiable, we only show the more complex 6.7-GHz maser morphology and present details of the 19.9-GHz maser in the figure caption. The size of the data markers are proportional to the flux densities of the maser components, and the error bars show the mean formal uncertainty of the spectral channels which have been averaged to give the position of the particular maser. The origin in each panel is the location of the peak spectral emission channel for that transition. The small deviations of the reference maser from the origin are the result of the flux-weighting process, which moves this peak off-centre when combined with emission at slightly offset positions. In the overlay panels, the origin is with respect to the peak 19.9-GHz maser emission and the data markers have not been scaled with flux density due to the very different intensities of the two transitions.

In the following sections dedicated to each individual source, we also discuss the spectra of the 19.9-GHz masers obtained in our observations (Figure \ref{fig:spectrum}) compared to those observed by \citet{Ellingsen+04}. The absolute flux density calibration is estimated to be better than 10 percent for the \citeauthor{Ellingsen+04} observations, with uncertainty in velocity being better than 0.1 \kms. We have optimised the panel sizes in Figures \ref{fig:w309} to \ref{fig:w353} to best display the distribution of maser emission for each source, so they are typically different for the different transitions (although always with a one-to-one aspect ratio).

\subsubsection*{G\,$309.921+0.479$}
This is the only source in the current study which was not observed by \citet{Ellingsen+04}.  It was detected in (to date unpublished) additional 19.9-GHz observations made with the Tidbinbilla 70m antenna, and these observations confirm that detection.  The current ATCA observations show a 19.9-GHz methanol maser with a peak flux density of 0.46~Jy, at a velocity of $-$60.9~\kms.  In comparison the 6.7-GHz transition has its strongest emission at a velocity of $-$59.7~\kms, and a weaker peak with a flux density of 86.7~Jy at essentially the same velocity ($-$60.8~\kms) as the 19.9-GHz maser (Figure~\ref{fig:spectrum}).

\begin{figure*}
\psfig{file=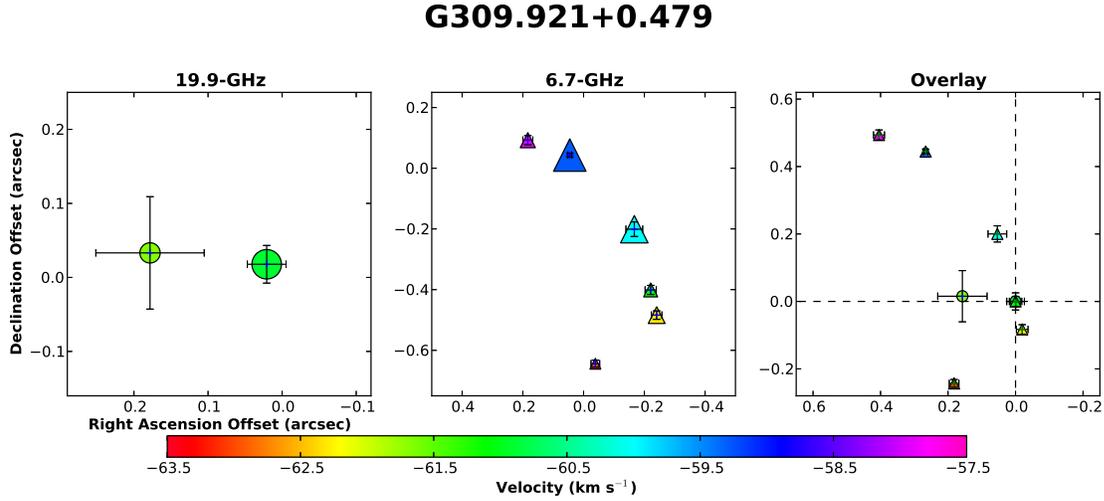,width=\linewidth , keepaspectratio=true}
\caption{Spatial distribution of the methanol maser components at 19.9- and 6.7-GHz. The 19.9-GHz components are at velocities of $-$61.7 and $-$60.9~\kms~with an east-west orientation and the velocity increasing from east to west. The 6.7-GHz emission has a curved profile. The spread of the components is mostly in the north-south direction and are spatially consistent with that observed by \citet{Phillips+98b}. The components have velocities of $-$63.0, $-$62.3, $-$61.1, $-$60.0, $-$59.3 and $-$57.9~\kms ~increasing from south to north. There is no evidence for spatial or velocity correlation between the masers at the two transitions.}
\label{fig:w309}
\end{figure*}

Figure~\ref{fig:w309} shows two 19.9-GHz maser components with an approximately east-west orientation and a separation of $<$0.2 arcsec. The spatial distribution of the 6.7-GHz masers lies roughly north-south, with maximum separation of 0.8 arcsec, similar to that observed by \citet{Phillips+98b}. The right-hand plot in Figure~\ref{fig:w309} shows that when the 19.9-GHz component at $-$60.9~\kms ~and the 6.7-GHz component at $-$61.1~\kms ~are aligned, there is no evidence for spatial or velocity correlation between the remaining components.

\subsubsection*{G\,$323.740-0.263$}
\begin{figure*}
\psfig{file=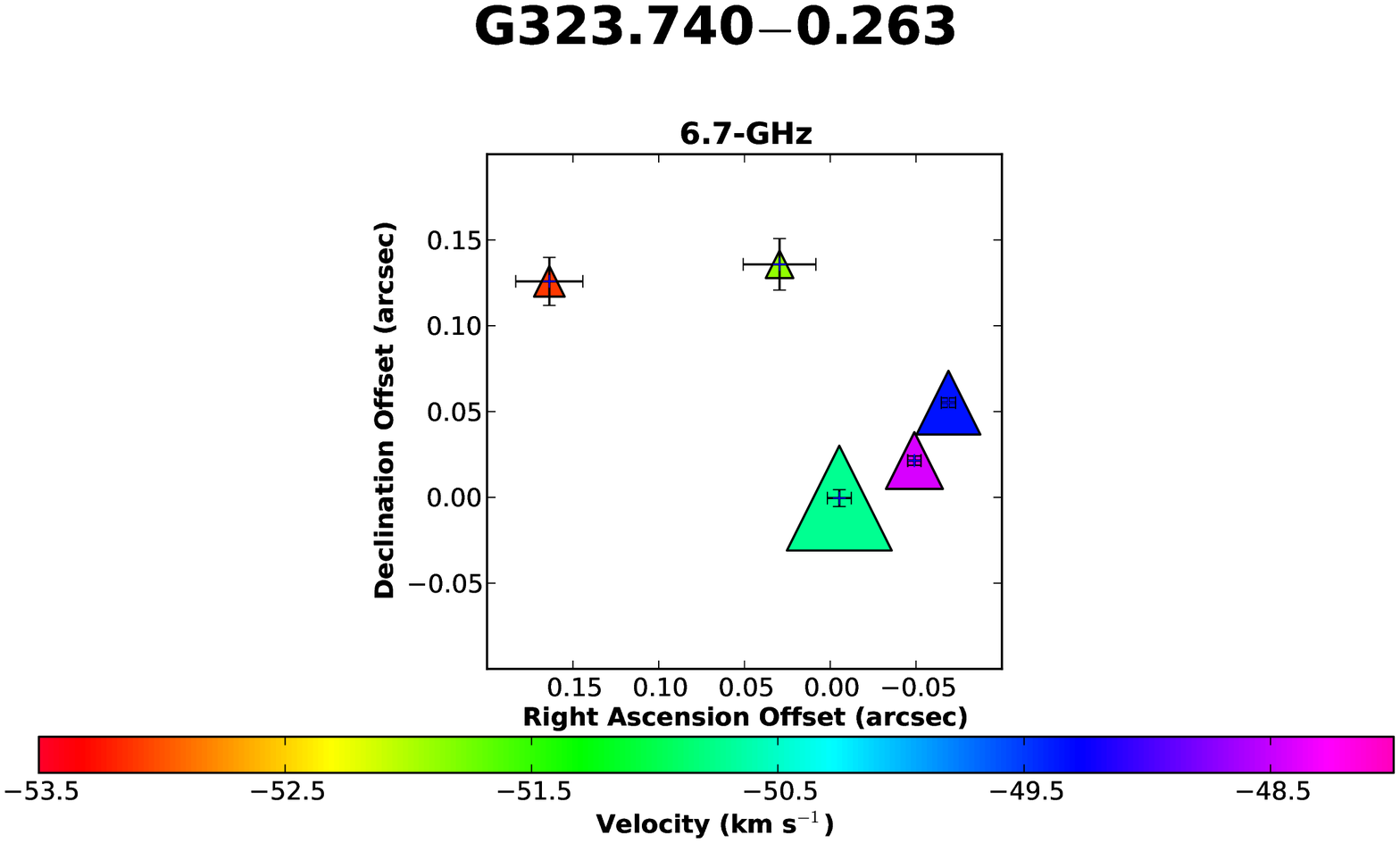, width=\linewidth , keepaspectratio=true}
\caption{The maser components at 6.7-GHz have a spatial distribution which is consistent with \citet{Norris+93, Walsh+02}. Velocity components were detected at $-$53.1, $-$51.9, $-$50.7, $-$49.3 and $-$48.4 ~\kms ~spanning $ 0.15 ^{\prime \prime }$ and $ 0.2 ^{\prime \prime }$ along the north-south and east-west axes respectively. The magnitude of the line of sight velocity generally increases from east to west, with the exception of the $-$48.4~\kms ~feature. Only a single 19.9-GHz maser component was identified at a velocity of $-$50.5~\kms and aligned with the $-$50.7~\kms\/ component at 6.7 GHz.}
\label{fig:w323}
\end{figure*}

\citet{Ellingsen+04} reported a peak flux density of 0.15~Jy at $-$51.4~\kms\/ for this source, the marginal detection being approximately 3 times the RMS noise level in their spectrum. The current observations (undertaken approximately a year and a half later) detected only very weak 19.9-GHz emission of 40~mJy, with an RMS of 7~mJy in the spectrally averaged (velocity resolution $\sim $1~\kms) image cube.  The initial detection by \citet{Ellingsen+04} of 19.9-GHz methanol emission in this source was with a relatively broad line width (3.1~\kms), possibly indicative of quasi-thermal emission, however, the large fractional change in flux density suggests that at least some of the emission must be from a maser, as we would not expect to see significant changes in thermal emission on such short timescales. Figure~\ref{fig:spectrum} shows that the major peaks of the 6.7-GHz spectrum, with flux density $\sim 2500$~Jy at $-$50.5 and $-$51.0~\kms ~coincide with the 19.9-GHz peak at this velocity resolution. Figure~\ref{fig:w323} shows the spatial distribution of the 6.7-GHz methanol maser emission, which agrees with the earlier images of \citet{Norris+93} and \citet{Walsh+02}. The morphology is curved with north-south and east-west separations of $\sim$0.2 arcsec.

\subsubsection*{G\,$328.808+0.633$}
\begin{figure*}
\psfig{file=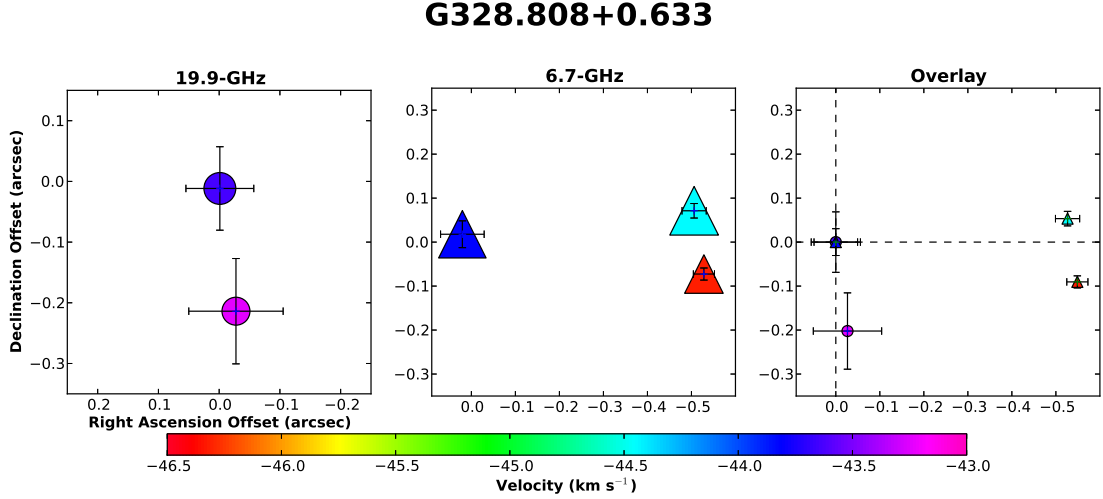, width=\linewidth , keepaspectratio=true}
\caption{Two 19.9-GHz methanol maser components were identified from the image cubes, at velocities of $-$43.2 and $-$43.6~\kms . The 6.7-GHz maser shows a predominantly east-west elongation of 0.5$^{\prime \prime}$ and are observed at velocities of $-$43.8~\kms\/ and $-$44.4, $-$46.3~\kms. We find no evidence for spatial or velocity correlation between the other 6.7- and 19.9-GHz components when we align the 19.9-GHz peak with the 6.7-GHz component with closest associated velocity.}
\label{fig:w328}
\end{figure*}

The 19.9-GHz methanol maser emission has a peak flux density of 0.63~Jy at $-$43.6~\kms.  Figure~\ref{fig:spectrum} shows that the 6.7-GHz methanol maser has its second strongest peak (flux density of 134~Jy) at essentially the same velocity ($-$43.7~\kms). \citet{Ellingsen+04} reported a peak flux density of 0.79~Jy at $-$43.7~\kms , with a FWHM of 0.9 \kms ~for the 19.9-GHz methanol maser. The current observations show the 19.9-GHz emission covering the same velocity range as the Tidbinbilla observations, with some evidence for weak, spectrally blended maser components on either side of the main peak. From our image cubes we are able to identify two maser components in the 19.9-GHz emission (Figure~\ref{fig:w328}). These are at velocities of $-$43.2 and $-$43.6~\kms ~and are separated by 0.2 arcsec, offset approximately north-south. This is quite different to the spatial distribution observed in the 6.7-GHz maser emission which shows elongation predominantly in the east-west direction, with a total angular separation of 0.5 arcsec between the component at $-$43.8~\kms ~and those at $-$44.4 and $-$46.3~\kms, consistent with \citet{Norris+93}. The right-hand panel in Figure~\ref{fig:w328} shows that when the 19.9- and 6.7-GHz components at $-$43.6~\kms ~and $-$43.8~\kms ~are aligned, there is no appreciable coincidence between the other components in terms of position or velocity.  The north-south separation between the maser components at 19.9~GHz and 6.7~GHz is similar, however, the velocity increases for the 19.9-GHz components but decreases for the 6.7-GHz masers, suggesting that any correlation is unlikely.

\subsubsection*{G\,$339.884-1.259$}
\begin{figure*}
\psfig{file=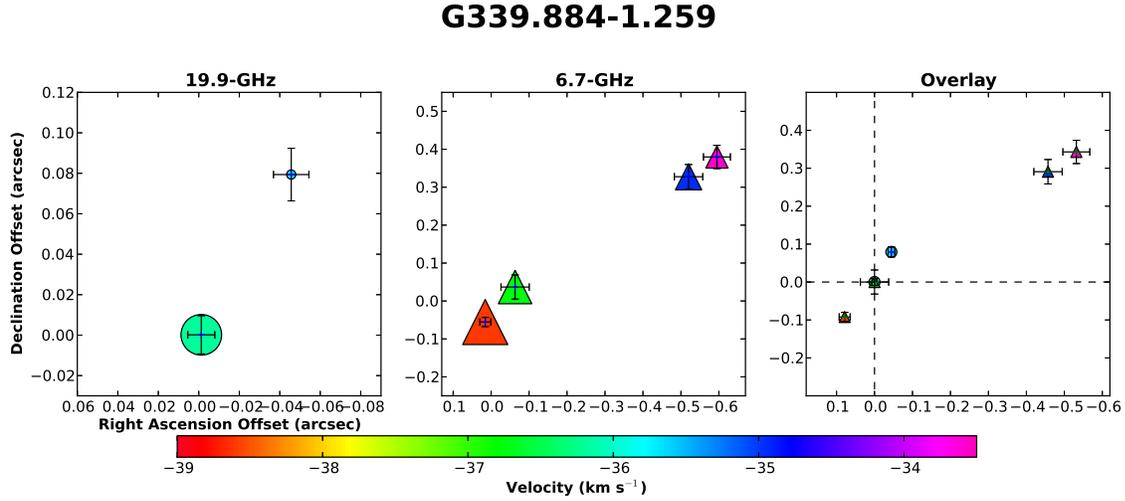, width=\linewidth , keepaspectratio=true}
\caption{Two 19.9-GHz methanol maser components were identified with velocities of $-$36.2 and $-$35.7~\kms. Four 6.7-GHz methanol maser components at velocities of $-$38.6, $-$36.7, $-$34.9 and $-$33.6~\kms ~were observed. Both transitions show a similar position angle for the maser distributions, and have velocity increasing in the same sense from east to west. However, no spatial or velocity correlation is evident.}
\label{fig:w339}
\end{figure*}

This is the second strongest 19.9-GHz methanol maser in our observed sample, with a peak flux density of  10.5~Jy at a velocity of $-$36.2~\kms.  This source also shows two additional weaker spectral features on either side of the strongest emission. The 6.7-GHz methanol masers cover a large velocity range with at least six peaks resolved with our coarse spectral resolution; the strongest having a flux density of 1250~Jy at a velocity of $-$38.6~\kms . The closest velocity 6.7-GHz maser component to the 19.9-GHz peak has a flux density of $\sim$500~Jy  at a velocity of $-$36.7~\kms\/ (Figure~\ref{fig:spectrum}). \citet{Ellingsen+04} report a 19.9-GHz peak flux density of 9.8~Jy at $-$36.2~\kms ~with FWHM of 0.4~\kms, consistent with the current observations. Looking at the relative distribution of maser components in the two transitions (Figure~\ref{fig:w339}), the 19.9-GHz emission traces a linear morphology in the north-west direction with an extent of $<$0.1 arcsec along either axis. The orientation is similar to that observed at 6.7~GHz \citep{Norris+93}, however, the 6.7-GHz methanol masers span a linear separation of 0.8 arcsec and there is no spatial or velocity coincidence between the two transitions.

\subsubsection*{G\,$345.003-0.224$}
\label{sec:subsec345}
The 19.9-GHz methanol maser has a single peak, with a flux density of 0.26~Jy at a velocity of $-$26.5~\kms\/ (Figure~\ref{fig:spectrum}). The 6.7-GHz maser spectrum shows many more peaks with the strongest having a flux density of 54.7~Jy at velocity $-$26.1~\kms, and evidence for blending with a secondary peak which is unresolved at our velocity resolution. This lies at the high-velocity extreme of the 19.9-GHz methanol maser velocity range. \citet{Caswell+10} show a spectrum where two peaks at 6.7~GHz are resolved at $-$26 and $-$27 \kms, meaning the 19.9-GHz maser peak lies between the two. \citet{Ellingsen+04} detected 19.9-GHz emission with a peak flux density of 0.22~Jy at a velocity of $-$27.4 kms (FWHM of 3.4~\kms). \citet{Ellingsen+04} reported the 19.9-GHz emission as being associated with G\,$345.003-0.223$, whereas our high spatial resolution ATCA observations show that the 19.9-GHz masers are associated with the nearby 6.7-GHz methanol maser G\,$345.003-0.224$ which is separated from G\,$345.003-0.223$ by 3.5 arcsec.  \citet{Caswell+10} report ``no significant overlap'' in the velocity range of these two neighbouring sources. Figure~\ref{fig:spectrum} shows that there is noticable overlap in some of the weaker emission in the velocity range between $-$29.8 and $-$22.8~\kms, and is consistent with \citet{Caswell97,Walsh+98}. The differences in these studies between the level of overlap is not explored in detail here, but can perhaps be attributed to changes in the morphology in the spectra, as a result of the range of epochs during which this source was observed.

\begin{figure*}
\psfig{file=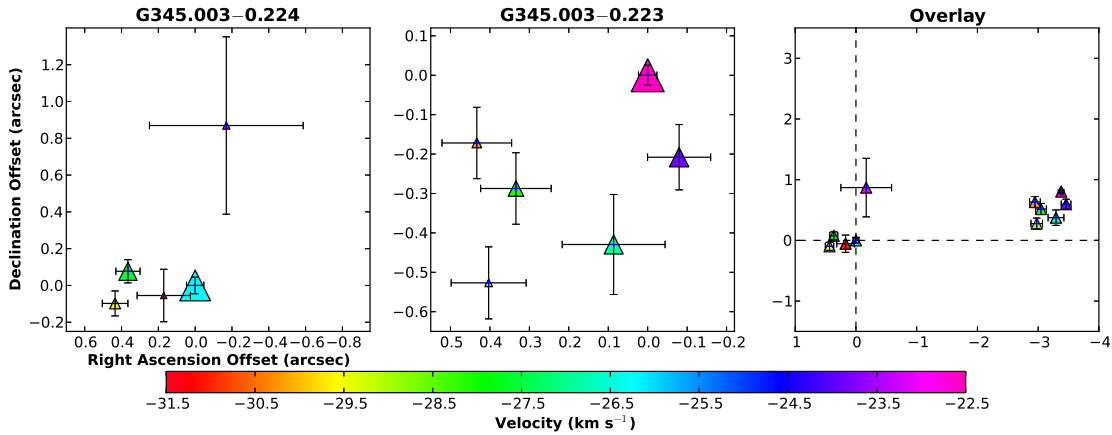, width=\linewidth , keepaspectratio=true}
\caption{6.7-GHz methanol masers associated with G\,$345.003-0.224$ (left cluster) range between $-$22.8 and $-$31.1 \kms ~LSR velocity, and G\,$345.003-0.223$ (right cluster) range between $-$22.6 and $-$30.2 \kms ~LSR velocity. In both sources, the LSR velocity increases from east to west except for the $-$31.1 \kms (weakest) feature associated with  G\,$345.003-0.224$. The overlay panel shows that the two sources have a separation of 3.5$^{\prime\prime}$ along an approximately east-west direction. The single 19.9-GHz methanol maser at $-$26.9~\kms\/ was identified to be associated with G\,$345.003-0.224$. It aligns closest to the $-$26.2 \kms \/ 6.7-GHz component associated with this source.}
\label{fig:w345}
\end{figure*}

Five maser components were identified from the 6.7-GHz image cubes of G\,$345.003-0.224$. The components have velocities ranging between $-$22.8 and $-$31.1~\kms, and are shown in the left panel in Figure~\ref{fig:w345}. The spread is predominantly in a north-south direction, spanning 1.0 arcsec, with less scatter ($\sim $0.5 arcsec) east-west. G\,$345.003-0.223$ is shown in the centre panel in Figure~\ref{fig:w345} with components ranging between $-22.6$ and $-30.2$~\kms. These components have a complex distribution spanning a range of $\sim$0.5 arcsec east-west. The spatial and kinematics distribution of the 6.7-GHz methanol maser emission for both sources are consistent with those presented in \citet{Caswell97}. The sources are separated by 3.5 arcsec along an east-west axis as shown in the right panel of Figure~\ref{fig:w345}. 

\subsubsection*{G\,$345.010+1.792$}
\begin{figure*}
\psfig{file=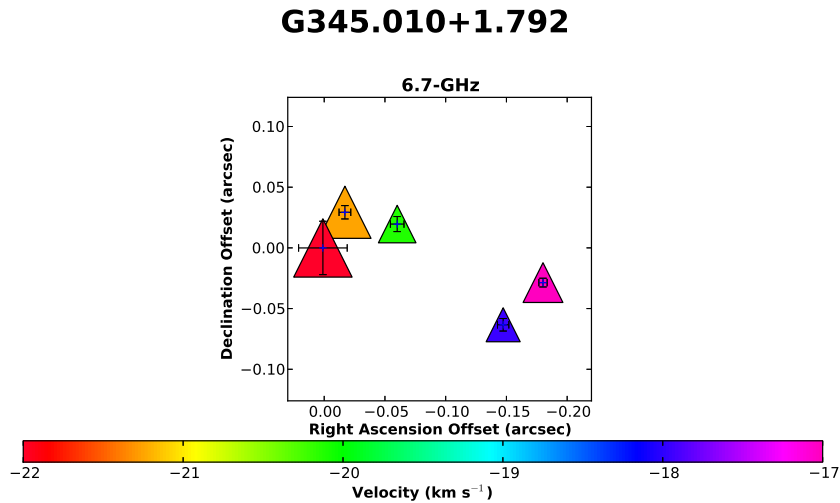, width=\linewidth , keepaspectratio=true}
\caption{A total of 5 maser components were identified from the velocity cube of the 6.7-GHz observations. These have velocities of $-$17.0, $-$17.9, $-$20.1, $-$21.2 and $-$22.3~\kms\/, with increasing LSR velocity from east to west. A single maser is identified at 19.9-GHz at a velocity of $-$17.5~\kms and aligned with the $-$17.9~\kms\/\ component at 6.7 GHz.}
\label{fig:w345010}
\end{figure*}

\citet{Ellingsen+04} detected a 19.9-GHz methanol maser with peak flux density of 0.21~Jy and at a velocity of $-$17.6~\kms\/ (FWHM of 1.5~\kms). The current ATCA observations show this maser to have a similar peak flux density of 0.25~Jy at $-$17.5~\kms.  The spectrum in Figure \ref{fig:spectrum} shows the velocity width of the maser feature to be less than 1~\kms. The 6.7-GHz spectrum has at least six spectral peaks with the strongest having a flux density of 218~Jy at $-$22.4 \kms . The 19.9-GHz methanol maser has a velocity which lies between two secondary 6.7-GHz maser peaks, with flux densities of 72.4 and 97.7~Jy and velocities of $-$18.0 and $-$16.9 \kms, respectively. G\,$345.010+1.79$ has been detected in a larger number of different class~II methanol maser transitions than any other source, including some that have only been detected in this source \citep{Ellingsen+12}, however, in general the rare, weak maser transitions are at velocities close to $-$22~\kms (the 19.9-GHz transition being the only exception).  The individual 6.7-GHz maser components identified from the image cubes are distributed into two distinct groups in Figure~\ref{fig:w345010}. The observed distribution is consistent with that shown by \citet{Norris+93}, covering an angular range of 0.2 arcsec east-west and a north-south spread of $<$0.1 arcsec.

\subsubsection*{NGC6334F}
\begin{figure*}
\psfig{file=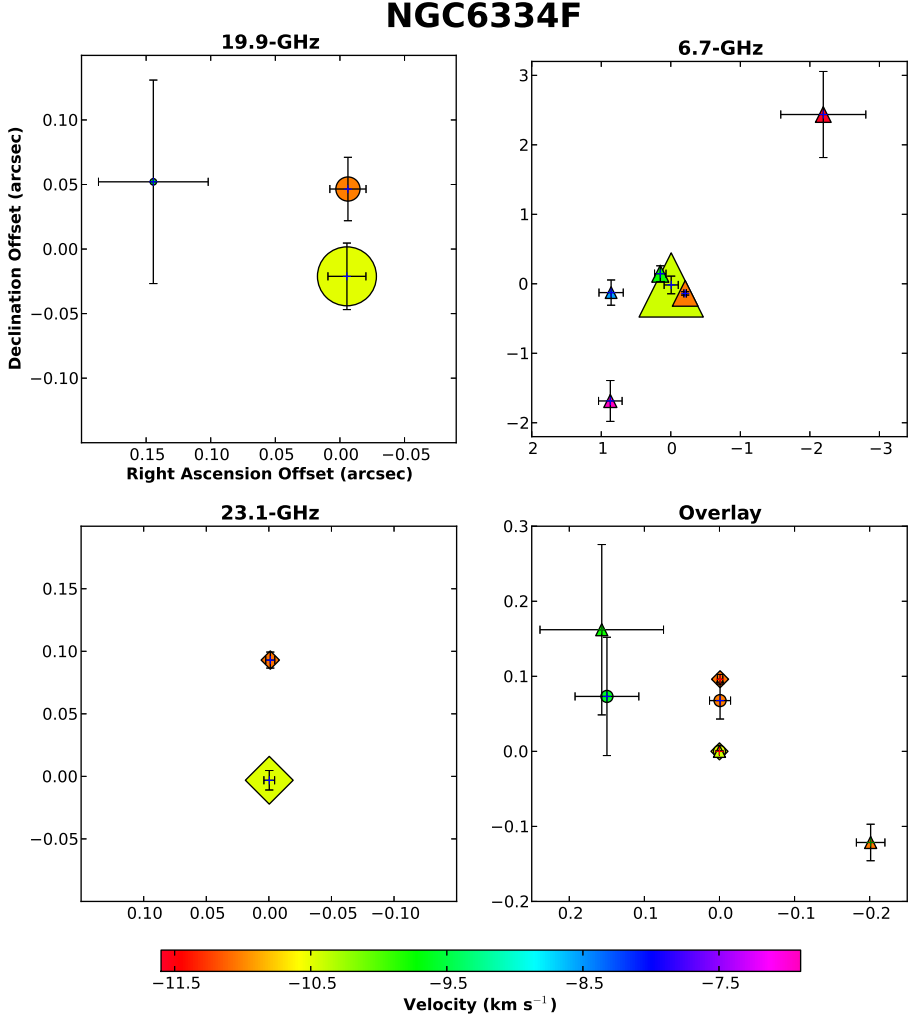, width=0.8\linewidth , keepaspectratio=true}
\caption{There are three maser components observed at 19.9-GHz at velocities of $-$9.6, $-$10.5 and $-$11.0~\kms. The two components at 23.1~GHz have velocities of $-$10.5 and $-$11.1~\kms. The bottom right-hand panel shows the primary maser cluster with components at 23.1-GHz aligning with corresponding 19.9-GHz emission to within 0.1$^{\prime \prime}$. The 6.7-GHz masers have velocities ranging between $-$11.6 and $-$6.9 \kms. There is possibly an alignment between the $-$9.6~\kms ~maser at 19.9 GHz and $-$9.8~\kms ~maser at 6.7 GHz as seen in the bottom right panel.}
\label{fig:wNGC}
\end{figure*}

This is the strongest of the 19.9-GHz methanol masers observed in the current ATCA observations. The spectra of the 6.7-, 19.9- and 23.1-GHz methanol masers are similar in their general structure, and individual peaks align closely in velocity (Figure~\ref{fig:spectrum}). The 19.9-GHz methanol maser has a peak flux density of 144~Jy at $-$10.5~\kms, similar to that observed by \citet{Ellingsen+04} who report a peak flux density of 146~Jy at $-$10.4~\kms ~and FWHM of 0.3~\kms. At 23.1~GHz the peak flux density (50.0 Jy) occurs at $-$10.5~\kms, while the 6.7-GHz maser peak (2230 Jy) occurs at a velocity of $-$10.3~\kms.  Figure~\ref{fig:wNGC} shows the 6.7-GHz methanol maser emission, with maximum separation of 4.0 arcsec between the components along a north-west elongation, consistent with \citet{Ellingsen+96a}.  The 19.9-GHz maser components have an east-west separation of $<$0.2 arcsec and a north-south separation of 0.1 arcsec. The bottom right panel of Figure~\ref{fig:wNGC} shows the region where the velocity components at all three transitions are close in proximity. The two maser components at 23.1~GHz ($-$10.5 and $-$11.1~\kms) are separated by 0.1 arcsec along the north-south axis, and align to within 0.1 arcsec (and 0.1~\kms\/ in velocity) with two of the three 19.9-GHz maser components (which are at velocities of $-$9.6, $-$10.5 and $-$11.0~\kms). The distribution of the 6.7-GHz emission is less certain due to the lower spatial resolution, although there is evidence for alignment between the 6.7-GHz masers and the 19.9-GHz component at $-$9.6~\kms\/, to within less than 0.1 arcsec.  Overall NGC6334F appears to be the most likely candidate for co-spatial emission of the 19.9-GHz methanol masers with other class~II transitions, especially given the similar spectra, however, VLBI observations are required to determine if spatially coincidence exists on milliarcsecond scales.

\subsubsection*{G\,$353.410-0.360$}
\begin{figure*}
\psfig{file=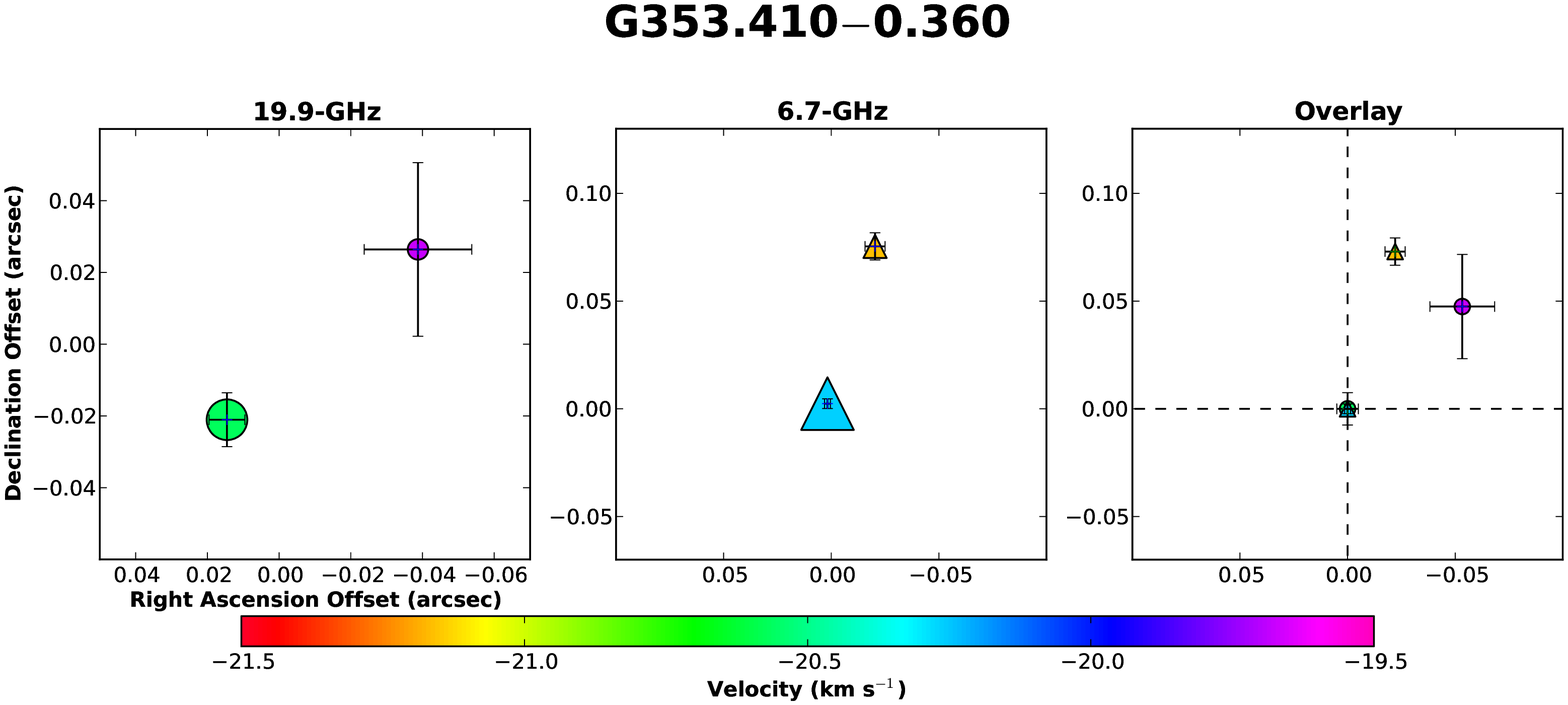, width=\linewidth , keepaspectratio=true}
\caption{The 19.9-GHz methanol maser emission has two components at velocities of $-$19.7 and $-$20.6 \kms. Our observations also show two maser components at 6.7~GHz at velocities of $-$20.3 and $-$21.2~\kms. No evidence for detailed spatial or velocity coincidence is apparent.}
\label{fig:w353}
\end{figure*}

We detect 19.9-GHz methanol maser emission with a peak flux density of 0.55~Jy at a velocity of $-$20.5~\kms ~in this source (Figure~\ref{fig:spectrum}). \citet{Ellingsen+04} measured the 19.9-GHz maser to have a peak flux density of 0.33~Jy at velocity $-$20.5~\kms ~with a FWHM of 1.8~\kms.  Our ATCA observations show secondary peaks on either side of the strongest emission which are likely why the Tidbinbilla observations required a relatively broad Gaussian component to fit the emission. At 6.7 GHz, G\,$353.410-0.360$ has two maser peaks detected at the spectral resolution of our ATCA observations (there is some indication of additional weaker emission at velocities similar to the strongest peak). The 6.7-GHz peak flux density is 89.6~Jy, at a velocity of $-$20.3~\kms. Two maser components were identified in the 19.9-GHz image cubes at velocities of $-$19.7 and $-$20.6~\kms (Figure \ref{fig:w353}).  The 6.7-GHz components have velocities of $-$20.3 and $-$21.2~\kms ~and emission in both transitions shows a similar position angle and separation of $<$0.1 arcsec.  The 6.7-GHz maser emission shows a distribution like that observed by \citet{Phillips+98b}. When the 19.9- and 6.7-GHz components at similar velocities are assumed coincident there is no evidence for spatial and velocity alignment of the other components (see right-hand panel of Figure~\ref{fig:w353}).

\subsection{Continuum}
\label{sec:maserContinuum}
\begin{figure*}
   \begin{center}
   \begin{minipage}[t]{0.4\textwidth}
       \psfig{file=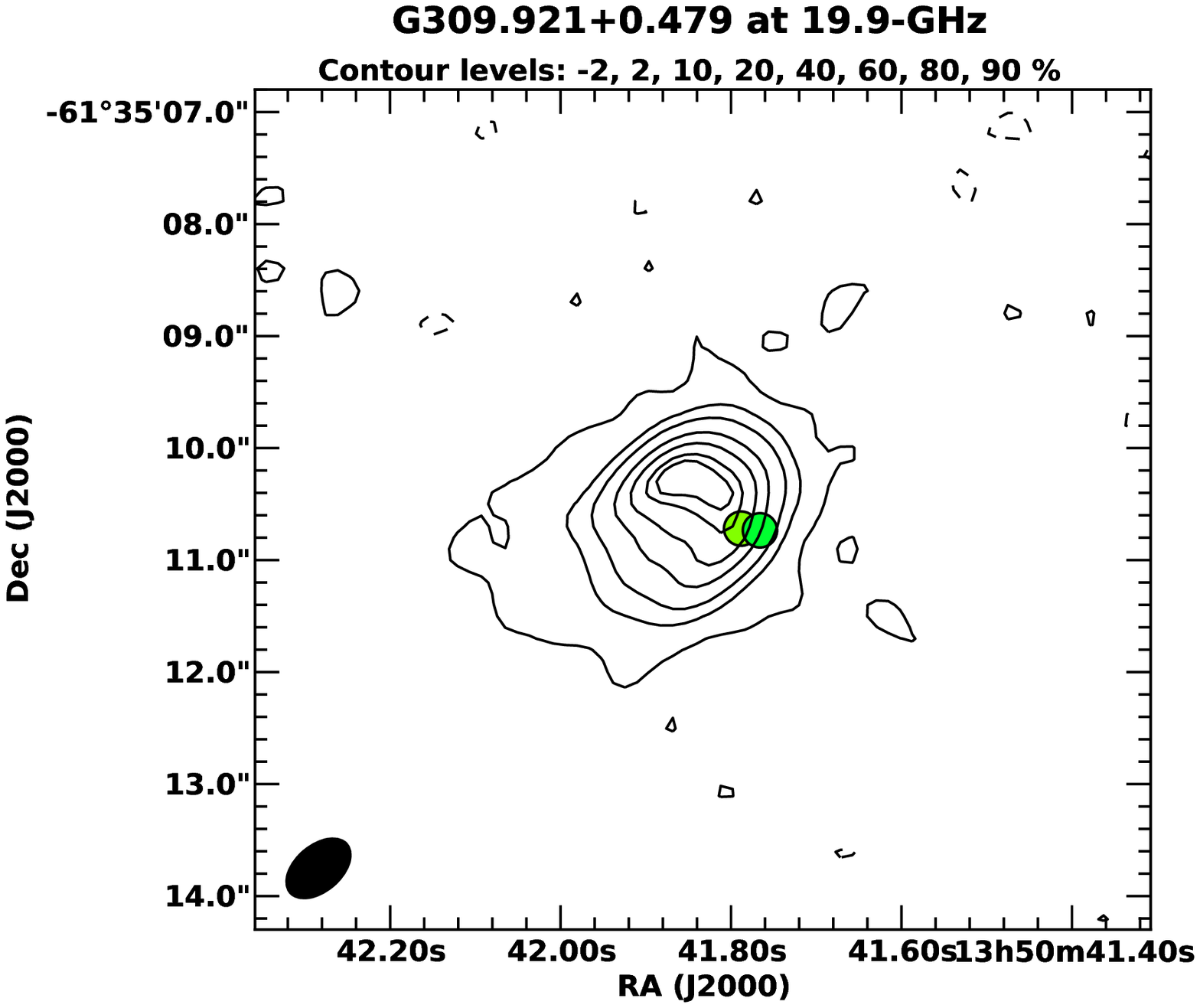,width=1.2\textwidth}
   \end{minipage}
   \begin{minipage}[t]{0.4\textwidth}
       \psfig{file=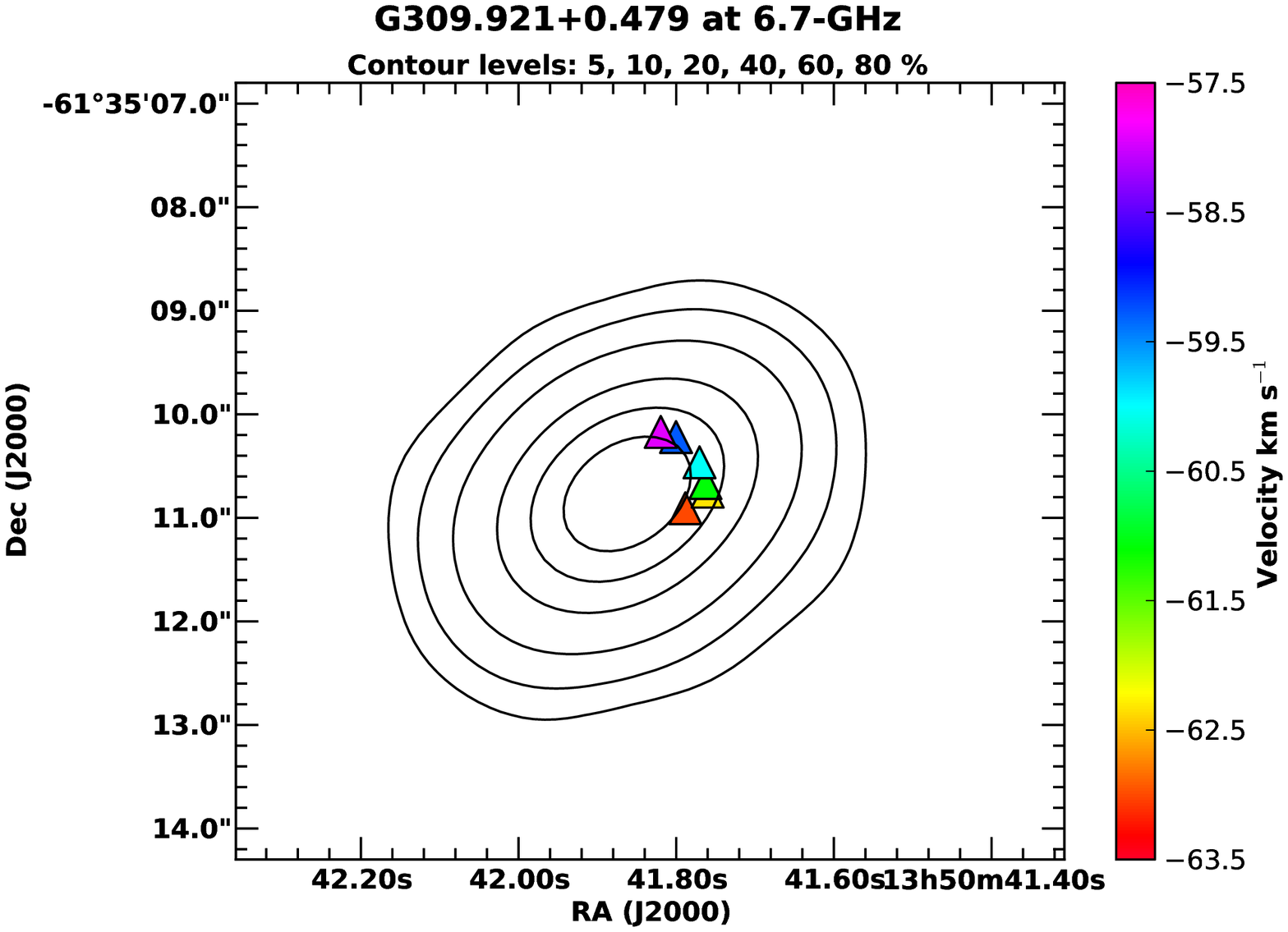,width=1.2\textwidth}
   \end{minipage}
 \end{center}
\caption{The 20-GHz continuum shows a cometary \ionhy region with the methanol maser emission ($-$61.7 and $-$60.9~\kms) projected against the south-west edge.  At 6.7~GHz the radio continuum emission is unresolved. The beam size at this frequency is $2.1 \times 1.4 ^{\prime\prime}$.}
  \label{fig:cont309}
\end{figure*}

\begin{figure*}
   \begin{center}
   \begin{minipage}[t]{0.4\textwidth}
       \psfig{file=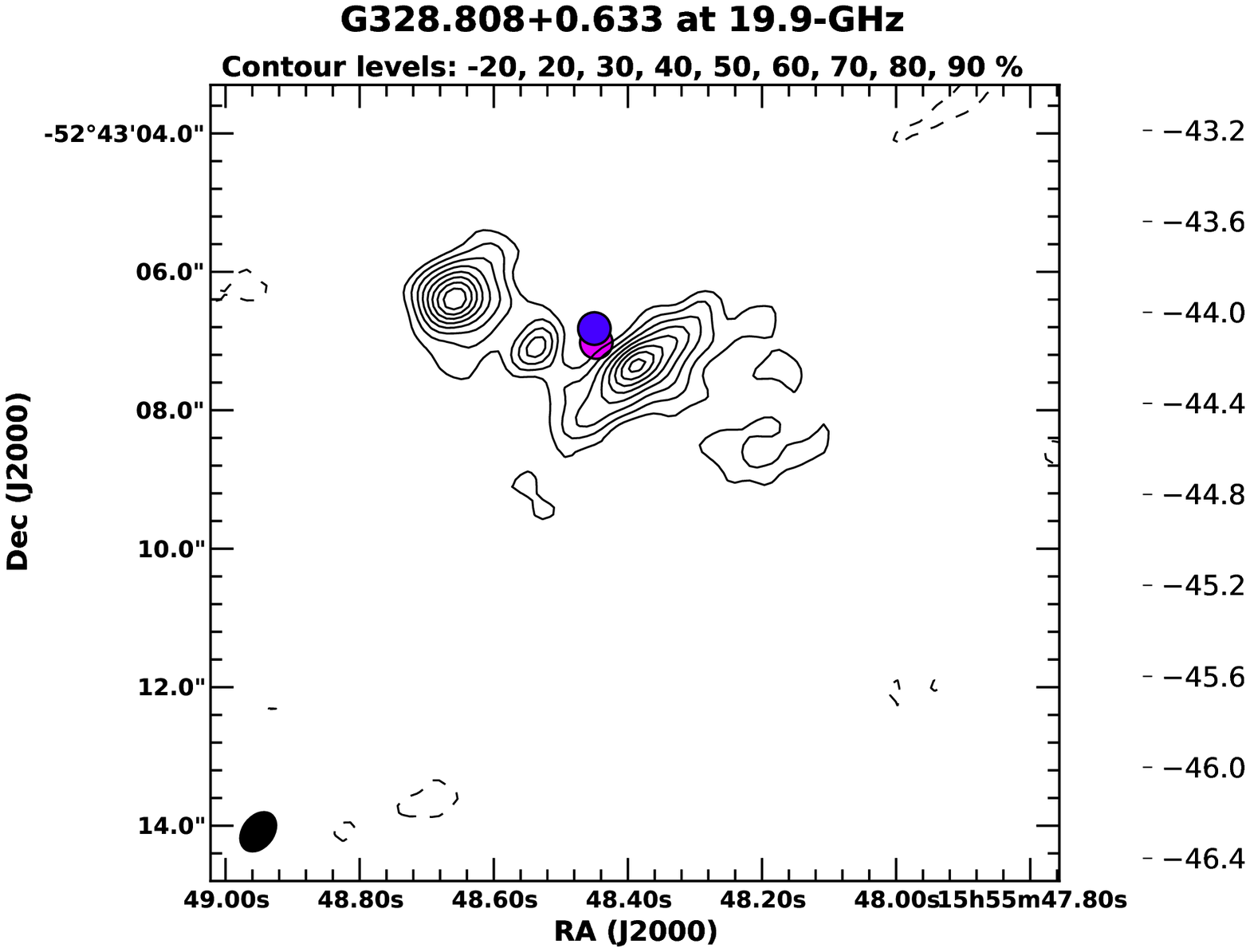,width=1.15\textwidth}
   \end{minipage}
   \begin{minipage}[t]{0.4\textwidth}
       \psfig{file=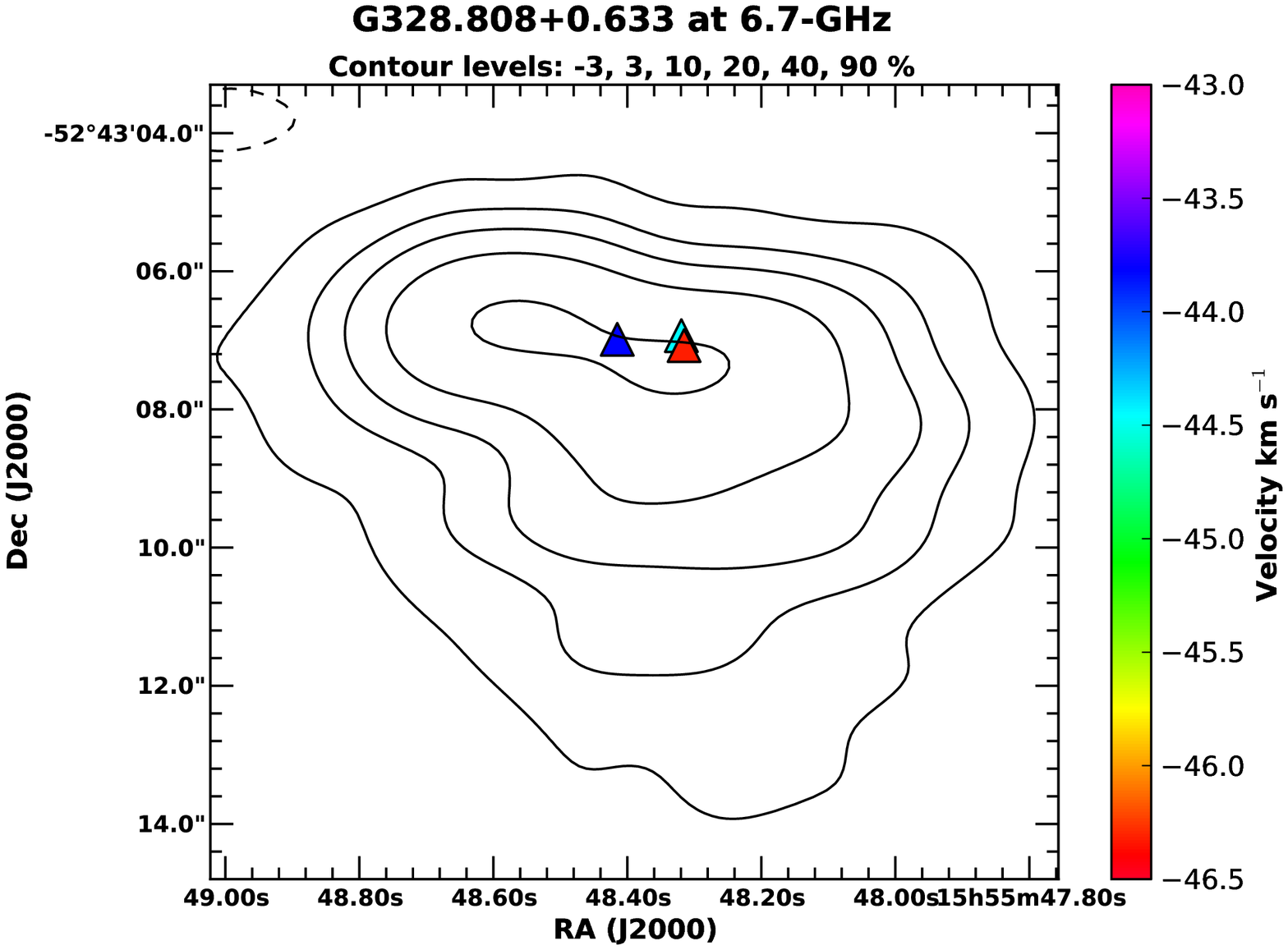,width=1.15\textwidth}
   \end{minipage}
 \end{center}
  \caption{The radio continuum at 20~GHz shows three compact components aligned roughly east-west. The masers at $-$43.2 and $-$43.6~\kms ~are not projected against radio continuum emission at this frequency, however, they do lie in regions where extended emission is observed at lower frequencies (e.g. the current 6.7-GHz observations and \citet{Ellingsen+05}). The beam size at 6.7 GHz is $2.5 \times 1.5 ^{\prime\prime}$.}
  \label{fig:cont328}
\end{figure*}

\begin{figure*}
   \begin{center}
   \begin{minipage}[t]{0.4\textwidth}
       \psfig{file=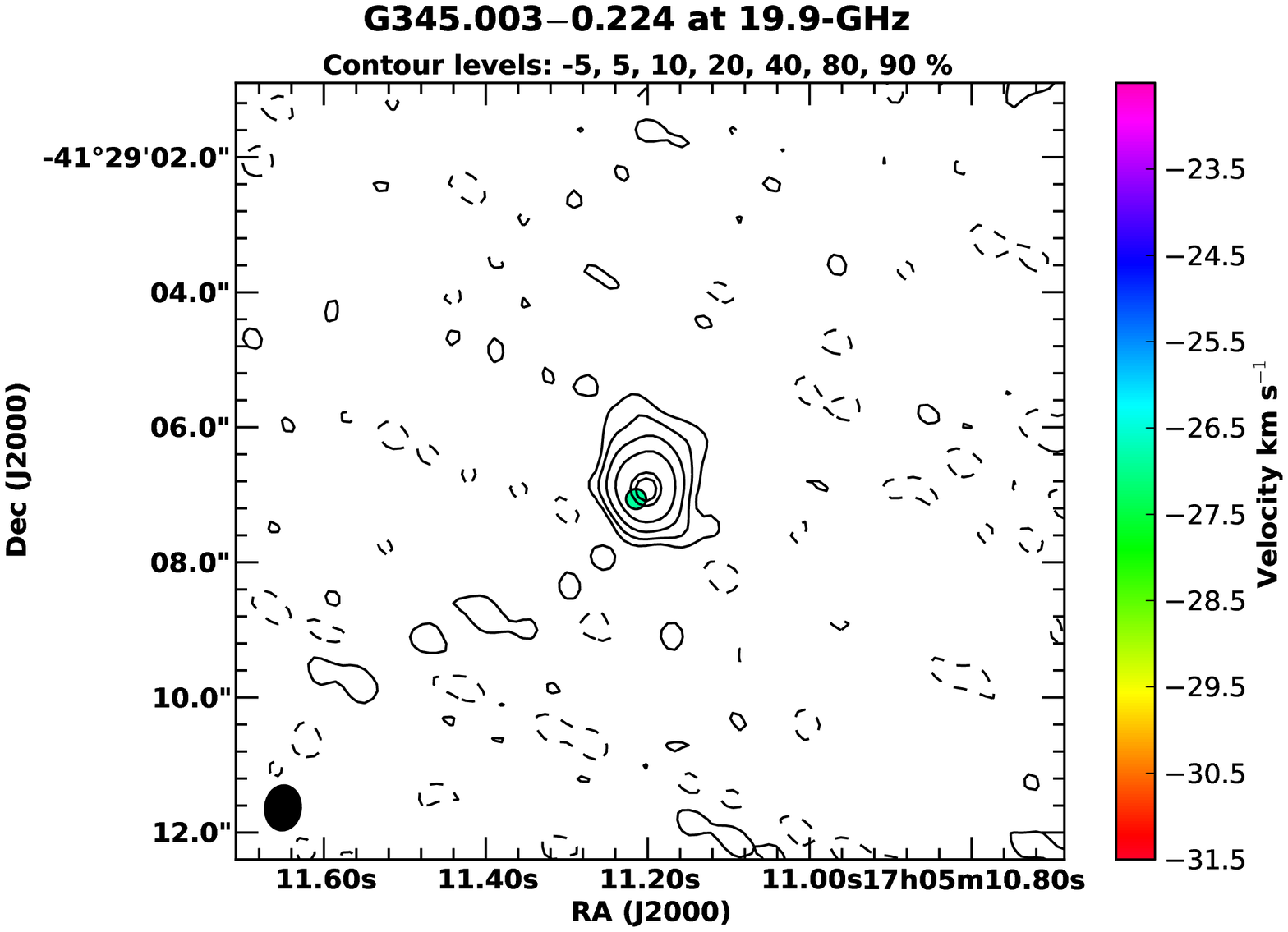,width=1.18\textwidth}
   \end{minipage}
   \begin{minipage}[t]{0.4\textwidth}
       \psfig{file=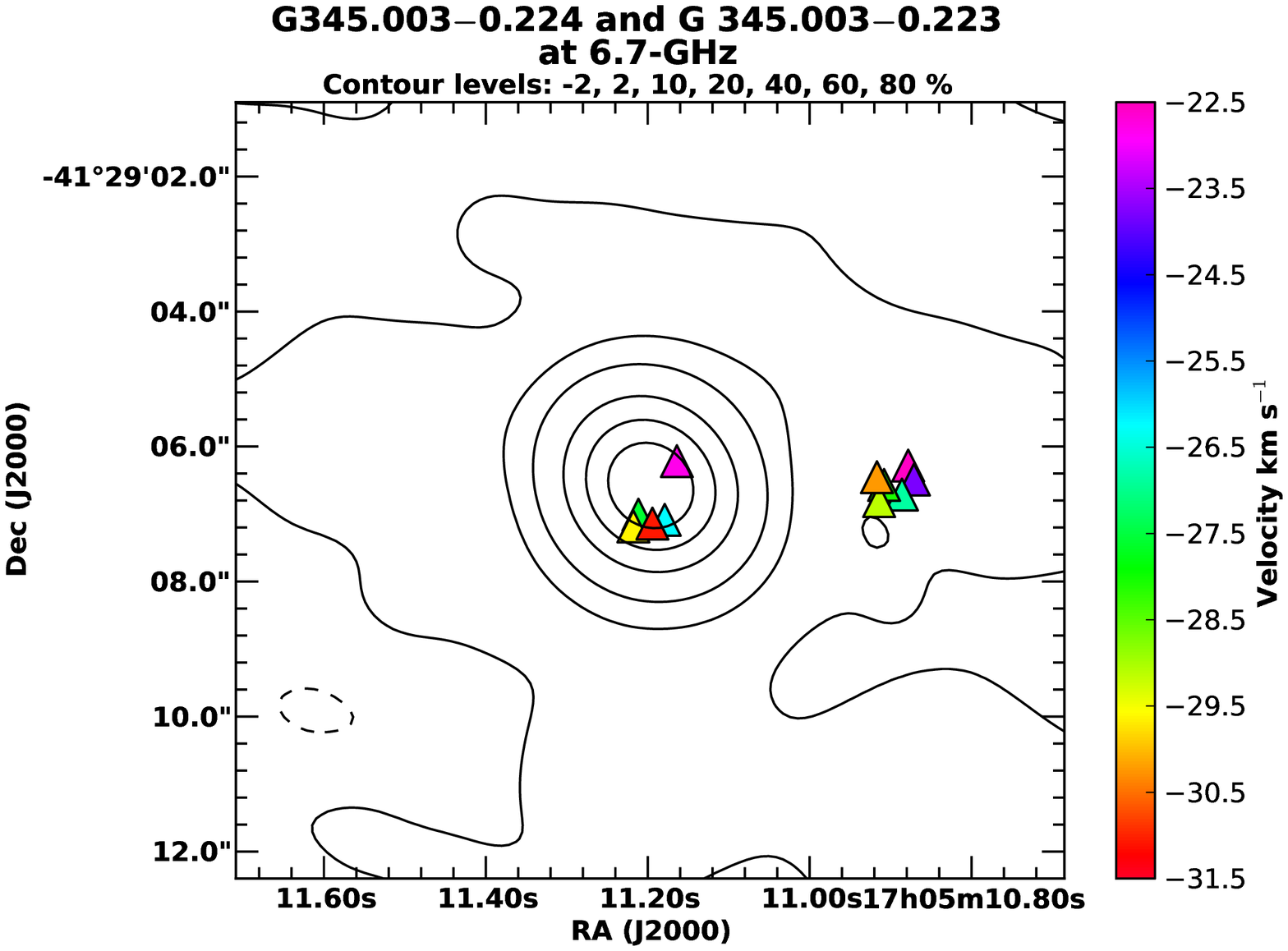,width=1.18\textwidth}
   \end{minipage}
 \end{center}
  \caption{The single maser component at 19.9~GHz ($-$26.9~\kms) is closely associated with radio continuum emission at this frequency. This is seen to be slightly resolved with a 0.5 arc second beam.  At 6.7~GHz, the two distinct groups of masers are associated with G\,$345.003-0.224$ (left cluster and closely associated with radio continuum) and G\,$345.003-0.223$ (right cluster and offset to the west of radio continuum). The beam size at 6.7 GHz is $2.1 \times 1.9 ^{\prime\prime}$.}
  \label{fig:cont345}
\end{figure*}

\begin{figure*}
   \begin{center}
   \begin{minipage}[t]{0.4\textwidth}
       \psfig{file=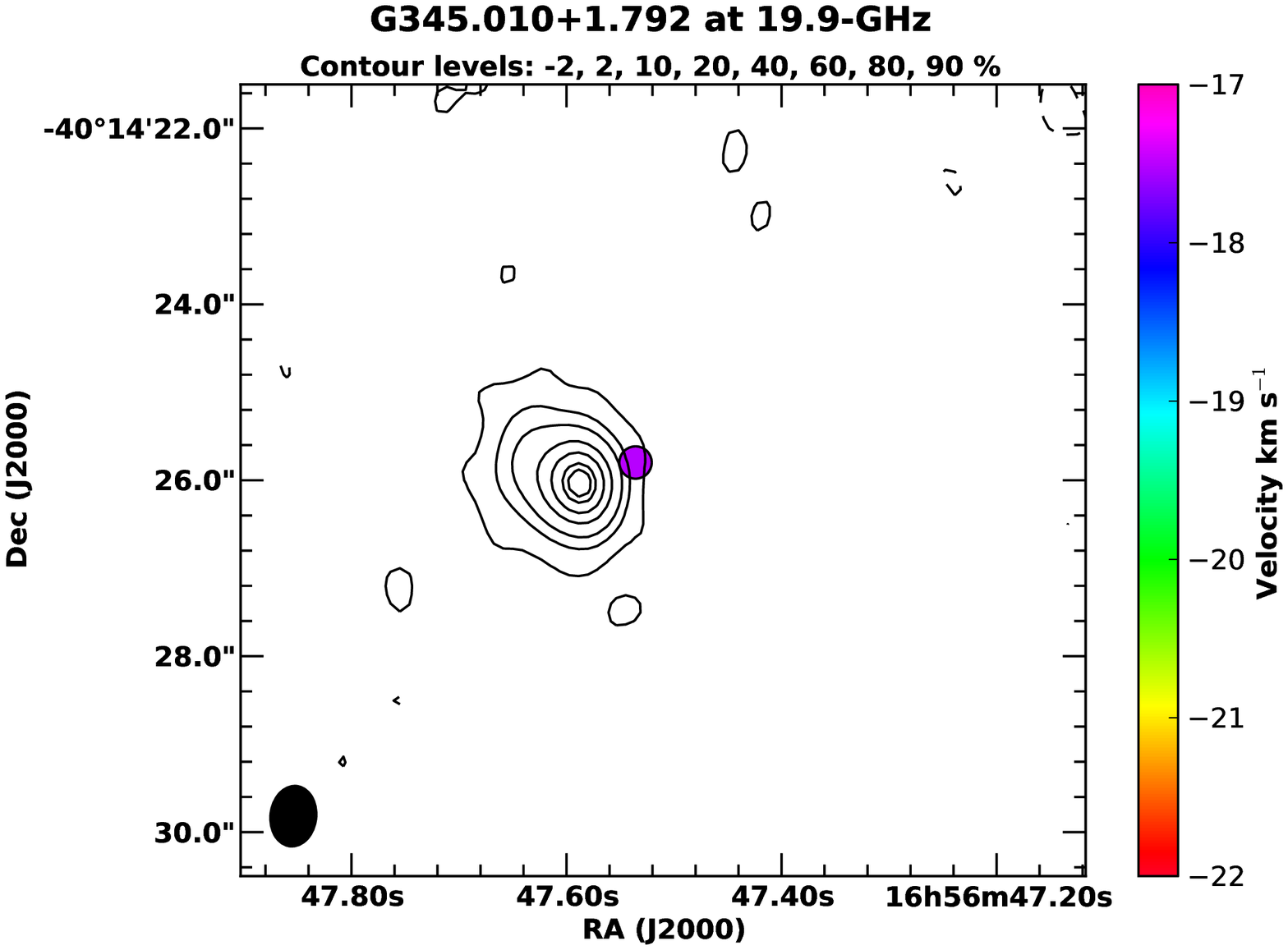,width=1.18\textwidth}
   \end{minipage}
   \begin{minipage}[t]{0.4\textwidth}
       \psfig{file=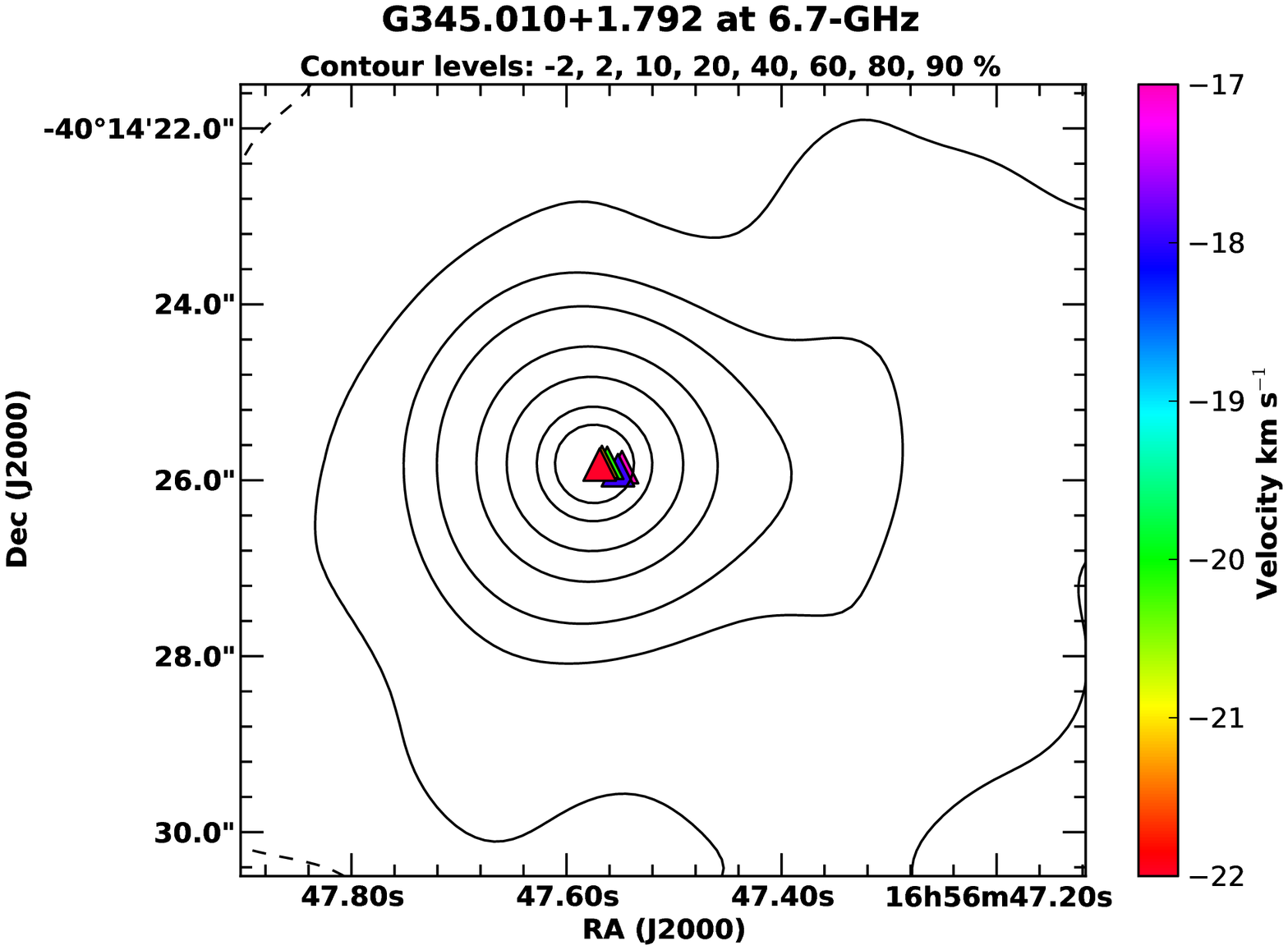,width=1.18\textwidth}
   \end{minipage}
 \end{center}
  \caption{At 19.9~GHz, the single maser at $-$17.5~\kms ~is observed to be slightly west of the continuum peak, which has slightly asymmetric profile, with the extended emission on the opposite side of the \ionhy region to the masers. The beam size at 6.7 GHz is $2.1 \times 1.9 ^{\prime\prime}$.}
  \label{fig:cont345010}
\end{figure*}

\begin{figure*}
   \begin{center}
   \begin{minipage}[t]{0.4\textwidth}
       \psfig{file=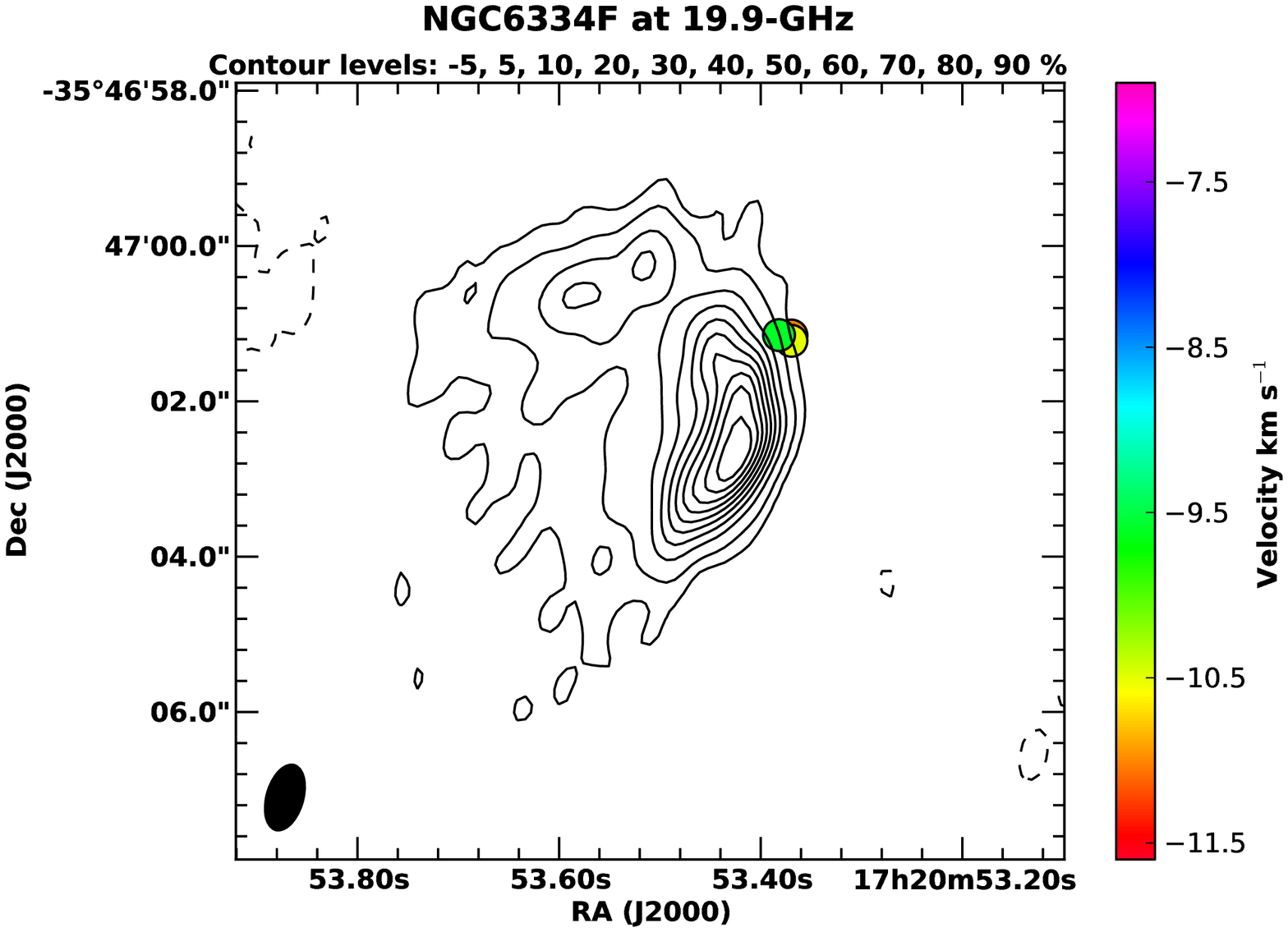,width=1.2\textwidth}
   \end{minipage}
   \begin{minipage}[t]{0.4\textwidth}
       \psfig{file=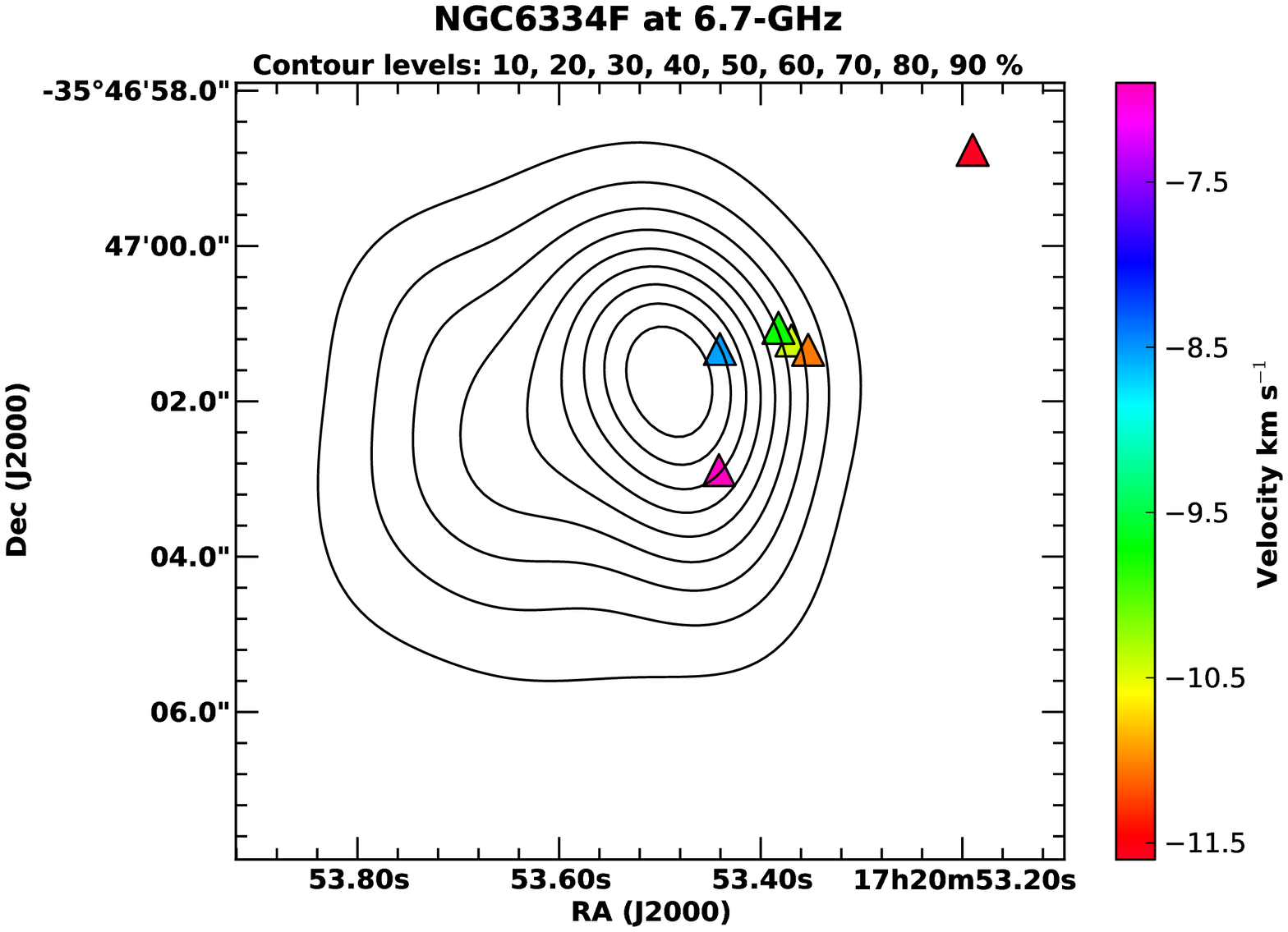,width=1.2\textwidth}
   \end{minipage}
   \begin{minipage}[t]{0.4\textwidth}
     \psfig{file=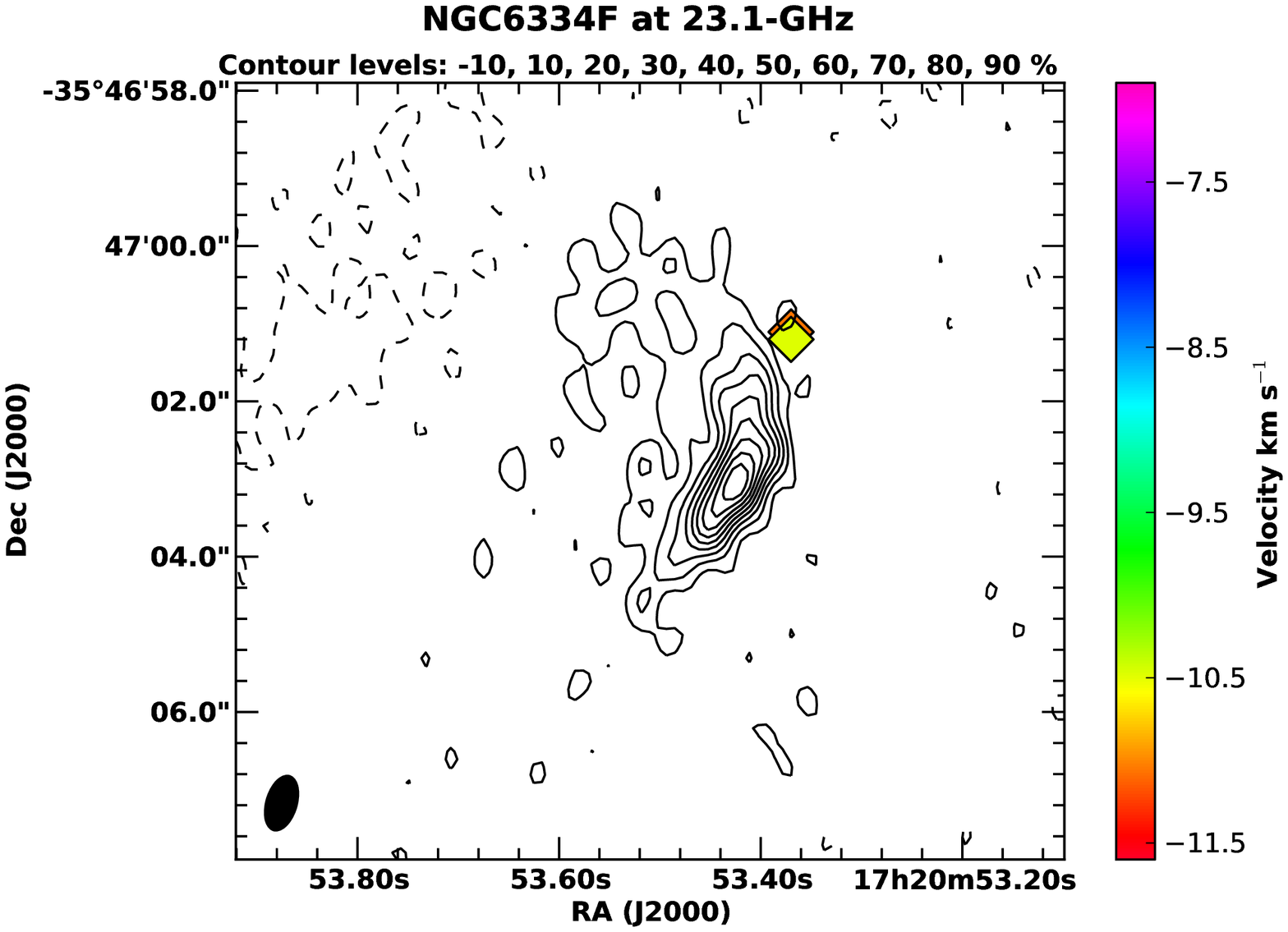,width=1.2\textwidth}
   \end{minipage}
  \end{center}
   \caption{The \ionhy region is cometary, with the 19.9-GHz maser emission at $-$9.6, $-$10.5 and $-$11.0~\kms\/  observed towards the leading edge.  At 6.7~GHz, additional maser clusters are observed both north-west and to the south of the cluster associated with the 19.9- and 23.1-GHz transitions. The beam size at 6.7-GHz is $2.5 \times 2.0^{\prime\prime}$.}
   \label{fig:contNGC}
\end{figure*}

\begin{figure*}
   \begin{center}
   \begin{minipage}[t]{0.4\textwidth}
       \psfig{file=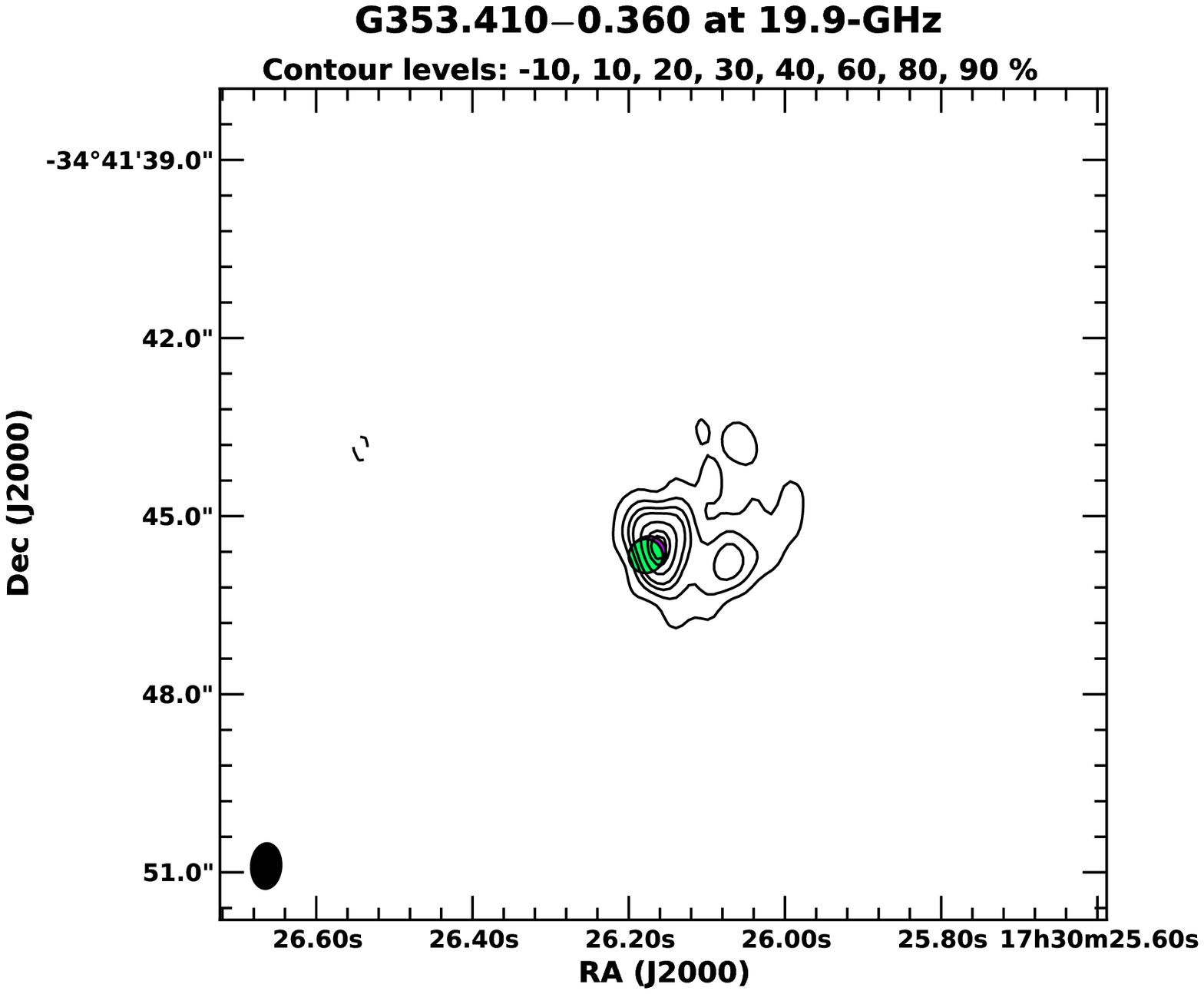,width=1.1\textwidth}
   \end{minipage}
   \begin{minipage}[t]{0.4\textwidth}
       \psfig{file=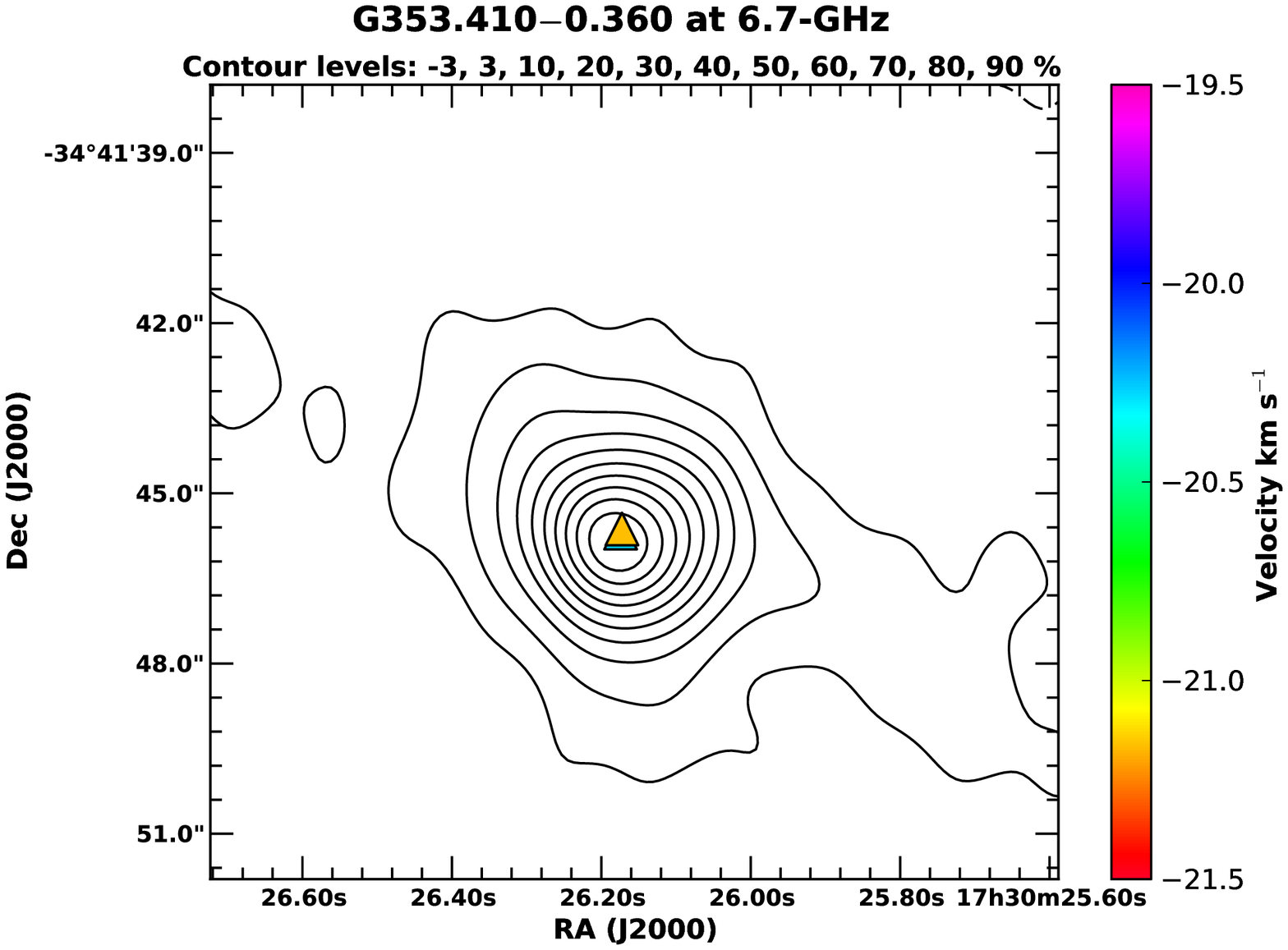,width=1.1\textwidth}
   \end{minipage}
 \end{center}
  \caption{The radio continuum emission for this source at 20~GHz has a cometary morphology with the masers at $-$19.7 and $-$20.6 \kms. They are observed to be projected against the leading edge of the \ionhy region. The masers at 6.7~GHz are also projected against the (unresolved) \ionhy continuum at this frequency. The beam size at 6.7 GHz is $2.4 \times 1.8 ^{\prime\prime}$.}
  \label{fig:cont353}
\end{figure*}

\citet{Ellingsen+04} noted that 19.9-GHz methanol masers are generally associated with 6.7-GHz methanol masers where the emission is projected against strong (greater than 100 mJy at 8.5~GHz) \ionhy continuum emission. Six of the eight 19.9-GHz methanol masers we observed are projected against radio continuum emission from an ultra-compact \ionhy region (see Section \ref{sec:ucHII} and Figures \ref{fig:cont309} to \ref{fig:cont353}). This is in contrast to the majority of known 6.7-GHz methanol masers which are not associated with radio continuum emission stronger than a few mJy \citep{Phillips+98b,Walsh+98,Minier+01,Bart+09,Pandian+10,Cyganowski+11}. The six \ionhy regions all have integrated flux density at 20~GHz stronger than approximately 80~mJy. The 19.9-GHz methanol masers for which we did not detect radio continuum emission were G\,$323.740-0.263$ and G\,$339.884-1.259$.  Sensitive ATCA observations by \citet{Ellingsen+96a} detected  an \ionhy region with an integrated flux density of 6.14~mJy at 8.6~GHz towards G\,$339.884-1.259$ (and also in \citet{Ellingsen+05}).  The spectral index of all the observed \ionhy\/ regions is less than 0.5 and even for a spectral index of unity, the expected 20-GHz integrated flux density for G\,$339.884-1.259$ is significantly less than the upper limit set by the current observations. Sensitive observations of G\,$323.740-0.263$ at 8.6~GHz by \citet{Walsh+02} failed to detect an \ionhy region associated with this source above 0.2~mJy, and in the light of this our non-detection at 20~GHz is expected for observations at the current sensitivity.

The properties of the 20-GHz radio continuum emission for the detections are summarised in Table~\ref{tab:contList}, and in the captions of Figures~\ref{fig:cont309} to \ref{fig:cont353}. In these figures, circles, triangles and diamonds represent 19.9-, 6.7- and 23.1-GHz masers respectively. The 20-GHz continuum observations have an approximate synthesised beam size of $\sim 0.5^{\prime \prime}$, compared to a 6.7-GHz beam of $\sim 2^{\prime \prime}$. Integrated fluxes were determined based on boxes which were similar in size to the \ionhy regions to mitigate effects due to different beam sizes between 6.7- and 19.9-GHz observations. The RMS for the peak and integrated fluxes were measured from a square box of $\sim10$ arcsec$^2$ in a region of the deconvolved image which was free from any continuum emission. These were found to be $\leq$20 mJy for the peak flux density and $\leq$40 mJy for the integrated flux density at 19.9 and 6.7 GHz. Detailed studies of these ultracompact \ionhy regions at 3.5~cm have been previously undertaken by \citet{Ellingsen+96a, Walsh+98, Phillips+98b}.

\begin{table*}
\centering
\caption{Properties of the radio continuum emission associated with 19.9-GHz methanol masers. The upper limit for sources without any detected continuum in the image is 5 times the RMS noise. The spectral indices, are based on a 2-point fit between the integrated flux densities at 19.9- and 6.7-GHz, and have errors of $<$0.1. The peak and integrated flux densities for NGC6334F for the 23.1-GHz observation are ($390 \pm 30$) and ($1900 \pm 60$)~mJy.}
\begin{tabular}{lrlrlrlrlcc}\hline
 \multicolumn{1}{c}{\bf Source} &   \multicolumn{8}{c}{\bf Flux Density}& \multicolumn{1}{c}{\bf Spectral}\\
 \multicolumn{1}{c}{\bf Name} & \multicolumn{4}{c}{\bf 19.9~GHz} & \multicolumn{4}{c}{\bf 6.7~GHz} &  \multicolumn{1}{c}{\bf Index}\\
 & \multicolumn{2}{c}{\bf Peak (mJy)}   &\multicolumn{2}{c}{\bf Integrated (Jy)}  & \multicolumn{2}{c}{\bf Peak (mJy)}   &\multicolumn{2}{c}{\bf Integrated (Jy)}  & \multicolumn{1}{c}{$\alpha$} \\
\hline
\hline 
 G\,$309.921+0.479$     & 197    & $\pm$ 3     & 872      & $\pm$ 7      & 380     & $\pm$ 5    & 480      & $\pm$ 10   & 0.5 \\
 G\,$323.740-0.263$      &          &                   &  $<$45 &                    &           &                    & $<$55  &                  &      \\
 G\,$328.808+0.633$     & 110    & $\pm$ 20   & 870       & $\pm$ 40    & 356     & $\pm$ 7    & 1410    & $\pm$ 10   & $-$0.4 \\
 G\,$339.884-1.259$      &          &                   & $<$45  &                    &           &                    & $<$35  &                  &      \\
 G\,$345.003-0.224$      & 180    & $\pm$ 10   & 290      & $\pm$ 20     & 162     & $\pm$ 5    & 200      & $\pm$ 10   & 0.3\\
 G\,$345.010+1.792$     & 182    & $\pm$ 3     & 345      & $\pm$ 6       & 179     & $\pm$ 5    & 230      & $\pm$ 10   & 0.4\\
 NGC6334F                     & 220    & $\pm$ 10   & 2030     & $\pm$ 20    & 830     & $\pm$ 20    & 2320    & $\pm$ 40   & $-$0.1 \\
 G\,$353.410-0.360$      & 84      & $\pm$ 4     & 349      & $\pm$ 8       & 271     & $\pm$ 8    & 490     & $\pm$ 20   & $-$0.3\\
\hline
\end{tabular}
  \label{tab:contList}
\end{table*}

\section{Discussion}
Our observations have made the first high spatial resolution investigation of the 19.9-GHz methanol maser transition.  Many of these sources are quite weak at 19.9-GHz with peak flux densities often $<$1~Jy.  This is in contrast to the 6.7-GHz methanol masers in the same regions.  The 19.9-GHz methanol masers typically have simple spectra with 1 or 2 identifiable components.  The maser emission at 19.9-GHz is found to be spatially coincident with 6.7-GHz maser emission to within approximately 0.2$^{\prime \prime}$, demonstrating that the two transitions arise from the same star formation site.  However, unlike comparisons of the relative distribution of 6.7- and 12.2-GHz methanol masers \citep[e.g.][]{Menten+92,Norris+93,Minier+00} there is little or no evidence for close correlation in the morphology of the 19.9- and 6.7-GHz masers.  The small number of maser features at 19.9~GHz and the difference in the spatial and velocity resolution between the 19.9- and 6.7-GHz observations makes such comparisons difficult. However, our results suggest that unlike the 12.2-GHz transition, the 19.9-GHz transition would not, in most sources, have a close correspondence (10's of milliarcsecond scale) with the 6.7-GHz methanol masers in the same region.

Figure~\ref{fig:spectrum} shows that the peak velocity of the 19.9-GHz maser aligns with the strongest emission in the 6.7-GHz maser spectrum for at most three sources (G\,$323.740-0.263$, NGC6334F  \& G\,$353.410-0.360$).  For some of the remaining sources (e.g. G\,$328.808+0.633$) the 19.9-GHz peak aligns with a secondary 6.7-GHz component. For other sources (e.g. G\,$339.884-1.259$ \& G\,$345.010+1.792$) there is no corresponding component in the 6.7-GHz emission identifiable from the spectrum.  \citet{Breen+11} found 80 percent of 12.2-GHz methanol masers show their strongest emission at the same velocity as the 6.7-GHz maser peak, while \citet{Ellingsen+13} found 37.7-GHz methanol masers to have the same peak velocity as the associated 12.2-GHz methanol masers in more than 50 percent of sources.  A close correspondence between the velocity of different transitions makes close spatial coincidence more likely and in contrast, it is unlikely for different maser transitions to be coincident if they have different line of sight velocities.  In this context, the lack of close spatial coincidence between 19.9-GHz methanol masers and the associated 6.7-GHz masers is not surprising, as the 19.9-GHz masers frequently have the velocity of their peak emission offset from that observed for the stronger transitions.

Multi-transition modelling, such as that undertaken by \citet{Sutton+01} and \citet{Cragg+01} is based on the assumption that the different maser transitions are co-spatial.  Clearly the evidence for the 19.9-GHz methanol masers suggests this is not likely to be the case for the majority of sources.  The degree to which the 19.9-GHz methanol masers are spatially coincident with other class~II maser transitions can only be rigorously determined through phase-referenced very long baseline interferometry experiments of the various transitions. We would expect such observations to show that the 19.9-GHz methanol maser emission generally arises in different locations within the larger class~II maser cluster from the stronger, more common transitions.  \citet{Ellingsen+04} found that 19.9-GHz methanol masers prefer lower gas and dust temperatures and higher densities than those sources which are not detected in this transition (e.g 19.9-GHz methanol masers prefer gas temperatures $T_k \sim$ 50~K, dust temperature $T_d \sim$ 175~K, density $n_H \sim 10^7$ cm$^{-3}$).  Given our results it would appear to be better not to include 19.9-GHz methanol masers in multi-transition models in sources where there is not a close alignment in velocity between the different transitions.

\subsection{Association with ultra-compact \ionhy regions}
\label{sec:ucHII}

\citet{Ellingsen+96a, Phillips+98b, Walsh+98} and others have shown that some 6.7-GHz methanol masers have an associated \ionhy region detectable at 8~GHz. The correlation between the observed 19.9-GHz methanol masers with strong radio continuum emission is indicative that these masers are associated with the later stages of the class~II methanol maser phase for young high-mass stars \citep{Breen+10a}.  For most of the \ionhy regions detected in our observations, previous observations with an angular resolution of around 2 arcseconds at 8.5~GHz showed them to be largely unresolved on these scales \citep{Ellingsen+05}.  The current observations at 20~GHz, with a resolution approximately a factor of 4 higher confirm the hints of asymmetry in the lower frequency radio continuum observations, and show that in all sources the 19.9-GHz methanol masers are projected against, or very close to the steepest gradient in the radio continuum emission.  Masers which are projected onto a \ionhy region have their brightness temperature boosted by seed photons  \citep{Sobolev+94}. As a result, it is possible that 19.9-GHz masers are found to be rare because they are low gain and only the sources which are projected against \ionhy regions are detected.

The majority of the \ionhy regions (5 of the 6 sources) have either a definite cometary morphology, or some indication of such a morphology. We take a cometary morphology to be one where there is an asymmetry in the intensity distribution with a well-defined axis consisting of closely spaced contours in the radio intensity on one side of a source and more diffuse emission on the opposite side.  We assess that the 19.9-GHz continuum images of all of the sources except G\,$328.808+0.633$ either clearly show, or indicate some signs of such morphology. In a large sample of \ionhy regions \citet{Wood+89a} found 20 percent of observed sources had a cometary morphology. Given the small sample size (and the subjective nature of morphological classification), we refrain from speculating whether 19.9-GHz methanol masers are preferentially associated with cometary \ionhy regions. However, if a larger sample becomes available it may be a question worthy of further investigation.

The spectral index ($\alpha $) of the radio continuum emission associated with each maser source was calculated using the integrated flux densities at 20.0~ and 6.7~GHz (see Table \ref{tab:contList}). These are all found to be $-0.5<\alpha<0.5$, confirming that the associated emission is thermal.  The calculated value of $\alpha = -0.4$ for G\,$328.808+0.633$ is likely to be due in part to the difference in spatial filtering between observations at different frequencies as seen in Figure \ref{fig:cont328}.  The measured spectral index suggests that the turnover between optically thick and optically thin free--free emission occurs at a frequency in between 7~and 20~GHz for these sources \citep{Hoare+07}.

\subsection{Maser-based evolutionary schemes}
The presence and absence of different maser transitions in different high-mass star formation region is indicative of differing physical conditions. \citet{Ellingsen+07} suggested that these differences could be used to characterise the evolutionary stage of high-mass star formation regions which have associated masers, and the proposed maser-based evolutionary timeline was subsequently quantified by \citet{Breen+10a}. In general, the strong, common maser transitions are most useful in such classifications, however, the rarer maser transitions must signpost regions with atypical conditions. This could be a star formation region with unusual properties, or perhaps more likely one which is in a relatively short-lived evolutionary phase. Recently \citet{Ellingsen+11a,Ellingsen+13} presented evidence that 37.7-GHz methanol masers are associated with star formation regions in the last few thousand years of the class~II methanol maser phase, and called these masers the ``Horsemen of the Apocalypse'' for this phase.  That the 19.9-GHz methanol masers do not in general appear to be cospatial with the more common class~II methanol maser transitions at the milliarcsecond level does not diminish their utility for evolutionary studies, as these focus on the large-scale physical properties of the broader region (i.e. comparable to, or larger than the scale of the maser clusters), where the transitions do coincide.

\begin{figure*}
\psfig{file=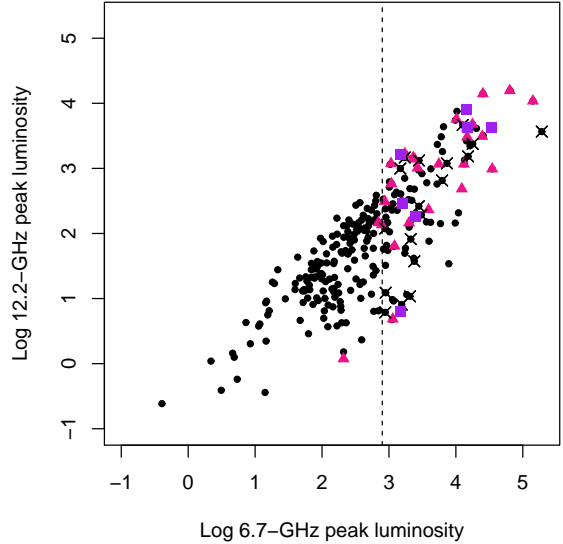, width=0.5\linewidth, angle=-90,  keepaspectratio=true}
\caption{The isotropic peak luminosity of the 6.7- and 12.2-GHz methanol maser emission towards all 12.2-GHz methanol masers south of declination -20$^\circ$ \citep{Breen+12a,Breen+12b}. The purple squares are sources with an associated 19.9-GHz methanol maser, the pink triangles are sources with an associated 107-GHz methanol maser (without an associated 19.9-GHz methanol maser). The dashed vertical line shows the 6.7-GHz peak luminosity cut off observed for 107-GHz methanol masers \citep{Ellingsen+11a}.  The black crosses indicate sources which have been searched for 107-GHz methanol maser emission, but were not detected.}
\label{fig:logLog}
\end{figure*}

Searches for the rarer, weaker class~II methanol maser transitions have generally been targeted towards star formation regions which show strong emission at 6.7 and 12.2~GHz.  The next most common class~II methanol maser transition is the 107-GHz ($3_1$ -- $4_0$A$^+$) transition, which has been detected towards 25 of more than 175 sites searched \citep{Valtts+95,Valtts+99,Caswell+00,Minier+02b}.  This sample of twenty five 107-GHz methanol masers has formed the basis of searches for the other rarer class~II methanol maser transitions over the last decade \citep[e.g.][]{Ellingsen+03,Cragg+04} and the results for a dozen transitions have been tabulated by \citet{Ellingsen+11a}.  The majority of the 107-GHz methanol maser sample are located in the southern hemisphere (22 sources), with the remaining three sources (W3(OH), Cep A and NGC7538) at high northern declinations.  The search for 19.9-GHz methanol masers by \citet{Ellingsen+04} targeted the 22 southern 107-GHz methanol maser sites. Combining their results with data from the literature for the three northern 107-GHz methanol masers \citet{Ellingsen+04} found that eight of the nine known 19.9-GHz methanol masers have an associated 6035-MHz excited-state OH maser (the exception being G\,$323.740-0.263$). This association implies that 19.9-GHz masers trace a specific range of physical conditions, likely linked with a late evolutionary phase \citep{Cragg+02}. \citet{Breen+10a} showed that the luminosity of both 6.7- and 12.2-GHz methanol masers increases as the source evolves, and \citet{Ellingsen+11a} showed that the sample of twenty five 107-GHz methanol masers typically have very high maser luminosities compared to the bulk of class~II masers. Figure~\ref{fig:logLog} plots the isotropic peak luminosity of the 6.7- and 12.2-GHz methanol masers towards all southern 12.2-GHz methanol maser sources \citep{Breen+12a,Breen+12b}, with all but one of the 107-GHz methanol maser sources lying to the right of the dashed line in this plot \citep{Breen+12b}, which are the higher 6.7-GHz luminosity sources. There are no clear trends for the properties of the 107-GHz methanol masers with and without an associated 19.9-GHz methanol maser, unlike the 37.7-GHz methanol masers, which are only associated with the higher luminosity 12.2-GHz methanol masers from the 107-GHz sample \citep{Ellingsen+11a}.

\citet{Ellingsen+11a} used the Onsala and Mopra telescopes to search for 37.7-GHz methanol masers towards all twenty five 107-GHz methanol maser sources. There are twelve 37.7-GHz methanol masers associated with the sample of twenty five 107-GHz masers, compared to nine 19.9-GHz methanol masers associated with the same 107-GHz sample. Five of the 107-GHz sample exhibit both 19.9- and 37.7-GHz class~II methanol masers (W3(OH), G\,$323.740-0.263$, G\,$339.884-1.259$, G\,$345.010+1.792$ and NGC6334F).  Of these five sources, four have an associated 6035-MHz OH maser (the exception being G\,$323.740-0.263$) and four have a peak 6.7-GHz flux density in excess of 1500~Jy (the exception being G\,$345.010+1.792$).  These five sources include all the strong (peak flux density $>$ 1500~Jy) 6.7-GHz maser sources, with the exception of G\,$9.621+0.196$.   In contrast, for the seven 37.7-GHz methanol maser sources which are not associated with a 19.9-GHz methanol maser, two have an associated 6035-MHz OH maser, and five do not.  

\citet{Caswell97} suggested the ratio of the 6.7-GHz to 1665-MHz OH peak flux density ($R$) is indicative of the age of a star formation region, with ``methanol-favoured'' sources ($R > 32$) being younger than those which are ``OH-favoured'' ($R < 8$).  Sixteen of the sample of twenty five 107-GHz methanol masers are associated with either or both of the 19.9- and 37.7-GHz transitions.  Table~\ref{tab:OHcomp} shows that there is a tendency for the 107-GHz methanol masers which have an associated 19.9-GHz methanol maser to have generally lower values of $R$, indicative of these being more evolved sources.  Although the sample size here is small and there is significant scatter for the observed methanol-OH peak flux density ratio, this trend appears consistent with the tendency for 19.9-GHz methanol masers to have an associated 6035-MHz OH maser and 37.7-GHz methanol masers not to, suggesting that the former trace a more evolved young high-mass star than the latter.

Based on the above trends, we suggest that the greatest number of class~II methanol maser transitions (including the 19.9- and 37.1-GHz transitions) are detectable close to the period when the 6.7-GHz masers reach their peak luminosity. Those sources where 37.7-GHz methanol masers are observed without an associated 19.9-GHz methanol maser likely precede the ``peak luminosity'' phase, while those with 19.9-GHz maser emission and no 37.7-GHz maser emission have evolved past ``peak luminosity''.  However, in making comparisons between the 19.9-GHz methanol masers and the 37.7-GHz methanol masers, it is important to note that the 19.9-GHz maser search of \citet{Ellingsen+04} is a factor of 5-10 more sensitive than the searches that have been undertaken for 37.7-GHz methanol masers (or any of the other rare maser transitions). Hence, it is possible (although not likely), that this difference in sensitivity may have indirectly introduced biases into the comparisons undertaken.

\begin{table*}
\centering
\caption{Comparison of the 6.7-GHz methanol maser to ground-state OH maser peak flux density ratio (R) for 107-GHz methanol maser sources with and without associated
 19.9-GHz and 37.7-GHz masers.}
\begin{tabular}{lcccc}\hline
\multicolumn{1}{c}{\bf Methanol Maser} & \multicolumn{1}{c}{\bf Methanol-favoured} &\multicolumn{1}{c}{\bf Intermediate} & \multicolumn{1}{c}{\bf OH-favoured} & \multicolumn{1}{c}{\bf Total} \\
\multicolumn{1}{c}{\bf Transitions} & \multicolumn{1}{c}{\bf ($R > 32$)} &\multicolumn{1}{c}{\bf ($8 < R < 32$)} &  \multicolumn{1}{c}{\bf ($R < 8$)} \\
\hline
\hline
37.7~GHz, no 19.9~GHz & 3 & 4 & 0 & 7\\
37.7~GHz \& 19.9~GHz   & 1 & 4 & 0 & 5 \\
no 37.7~GHz, 19.9~GHz  & 0 & 2 & 2 & 4 \\
\hline
\end{tabular}
  \label{tab:OHcomp}
\end{table*}

\section{Conclusions}
19.9-GHz methanol masers arise in high--mass star formation regions which also have associated 6.7- and 12.2-GHz methanol masers.  The 19.9-GHz methanol masers are coincident with the stronger, more common class~II methanol maser transitions to within 0.2$^{\prime\prime}$ in those sources where they are detected. Despite the similarity in the relative distribution of the 6.7- and 12.2-GHz methanol masers, both the velocity and spatial distribution of 19.9-GHz methanol masers suggests that they generally arise at different locations within the maser cluster.

Six of the eight sources are also found to be associated with strong ultra-compact \ionhy regions ($>80$mJy at 20 GHz) with five out of these exhibiting a cometary type morphology. This close association with strong ultra-compact \ionhy regions is in contrast to the majority of known 6.7-GHz methanol masers which are generally known not to be associated with radio continuum emission stronger than a few mJy.

We find that 19.9-GHz methanol masers are preferentially associated with OH masers with greater peak flux density (relative to the peak flux density of the 6.7-GHz methanol masers) than the 37.7-GHz methanol masers. Based on this relationship, while keeping in mind the small pool of 19.9-GHz masers in this study, we suggest that these masers likely trace more evolved high-mass star formation regions than those sources with an associated 37.7-GHz methanol maser.

\section*{Acknowledgements}
The Australia Telescope Compact Array is part of the Australia Telescope which is funded by the Commonwealth of Australia for operation as a National Facility managed by CSIRO.  This research has made use of NASA's Astrophysics Data System Abstract Service.  We would like to thank Patrick Ramsdale for his assistance with the ATCA observations.

\bibliography{maser_ref}

\end{document}